\documentclass[lettersize,journal]{IEEEtran}
\usepackage{amsmath,amsfonts}
\usepackage{algorithm}
\usepackage{array}
\usepackage[caption=false,font=normalsize,labelfont=sf,textfont=sf]{subfig}
\usepackage{textcomp}
\usepackage{stfloats}
\usepackage{url}
\usepackage{verbatim}
\usepackage{graphicx}
\usepackage{cite}

\usepackage{filecontents}

\usepackage{booktabs}
\usepackage{amsfonts}
\usepackage{multirow}
\usepackage{xcolor}
\usepackage{pifont}
\usepackage{algpseudocode}
\usepackage{colortbl}
\definecolor{darkblue}{HTML}{1D305F}
\usepackage{verbatim}
\usepackage{listings}
\usepackage{enumitem}
\usepackage{hyperref}
\hypersetup{hidelinks}
\usepackage{pifont}
\hypersetup{
	colorlinks=true,
	citecolor=red,
	linkcolor=green,
}

\usepackage[most]{tcolorbox}

\begin{document}

\title{Make Identity Unextractable yet Perceptible: Synthesis-Based Privacy Protection for Subject Faces  in  Photos}

\author{Tao Wang, Yushu Zhang, Xiangli Xiao, Kun Xu, Lin Yuan, Wenying Wen, and Yuming Fang

\thanks{Tao Wang and Kun Xu are with the College of Computer Science and Technology,
Nanjing University of Aeronautics and Astronautics, Nanjing 211106, China
(e-mail: \{wangtao21, xukun930\}@nuaa.edu.cn.)}
\thanks{Yushu Zhang, Wenying Wen, Xiangli Xiao, and Yuming Fang are with the School of Computing and Artificial Intelligence, Jiangxi University of Finance and Economics, Nanchang 330032, China (e-mail: zhangyushu@jxufe.edu.cn,  xiaoxiangli@jxufe.edu.cn;  wenyingwen@sina.cn; fa0001ng@e.ntu.edu.sg).}

\thanks{ Lin Yuan is with Chongqing Key Laboratory of Image Cognition, Chongqing University of Posts and Telecommunications, Chongqing 400065, China (e-mail:yuanlin@cqupt.edu.cn).}

}

\markboth{Journal of \LaTeX\ Class Files,~Vol.~14, No.~8, August~2021}%
{Shell \MakeLowercase{\textit{et al.}}: A Sample Article Using IEEEtran.cls for IEEE Journals}


\maketitle

\begin{abstract}
Deep learning-based face recognition (FR) technology exacerbates privacy concerns in photo sharing.  In response, the research community developed a suite of anti-FR methods to block identity extraction by unauthorized FR systems. Benefiting from quasi-imperceptible alteration, perturbation-based methods are well-suited for privacy protection of \textit{subject faces} in photos, as they allow familiar persons to recognize subjects via naked eyes. However,  we reveal that perturbation-based methods provide \textit{a false sense of privacy} through theoretical analysis and experimental validation. Therefore, new alternative solutions should be found to protect subject faces.

In this paper,  we explore synthesis-based methods as a promising solution, whose challenge is to enable familiar persons to recognize subjects. To solve the challenge,  we present a key insight—in most photo sharing scenarios, familiar persons recognize subjects through \textit{identity perception} rather than \textit{meticulous face analysis}.  Based on the insight, we propose the first synthesis-based method dedicated to subject faces, i.e., PerceptFace, which can make identity unextractable yet perceptible.  To enhance identity perception,  a new perceptual similarity loss is designed for faces,  reducing the alteration in regions of high sensitivity to human vision.

As a synthesis-based method, PerceptFace can inherently provide reliable \textit{identity protection}. Meanwhile, out of the confine of \textit{meticulous face analysis}, PerceptFace focuses on \textit{identity perception}  from a more practical scenario, which is also enhanced by the designed perceptual similarity loss.  Sufficient experiments show that PerceptFace achieves a superior trade-off between identity protection and identity perception compared to existing methods. We provide a public API  of PerceptFace and believe that it has great potential to become a practical anti-FR tool. Our code is available at https://github.com/daizigege/PerceptFace.
\end{abstract}.   


\section{Introduction}
\IEEEPARstart{P}{hoto}  sharing on online social networks (OSNs) \cite{beldad2017more} is a crucial service, that has revolutionized the way people record lives and express emotions. Nevertheless, with the development of deep learning-based face recognition (FR) technology, such a service also raises privacy concerns, which draws major attention from the research community and data protection agencies  \cite{10.1145/3673224}. The faces in the shared photos can be captured by individuals or organizations to train FR models,  automatically identified by FR systems for malicious purposes, or misused by DeepFake technologies \cite{verdoliva2020media} to enable face forgery.   In addition, sensitive attributes exposed in identity can cause additional potential harm to individuals \cite{9163294,li2023privacy}, e.g., sex discrimination. Notoriously, Clearview crawled more than 3 billion images from OSNs without the consent or knowledge of individuals to build its large-scale FR system. For payment, anyone is permitted to monitor people by its FR system.

In response, the research community has developed a number of anti-FR methods \cite{wenger2023sok,meden2021privacy} for face privacy.  Before uploading a photo to OSNs, users can process the faces in the photo with anti-FR methods, preventing the correct recognition by \textit{unauthorized FR systems}. Leaving aside methods (e.g., encryption \cite{ tajik2019balancing,yuan2015privacy} or transformation \cite{jin2024faceobfuscator,zhu2023campro,zhang2024validating}) that compromise visual quality,  existing anti-FR methods can be mainly categorized into two groups:  \textit{1) } Synthesis-based methods generate a face with a new identity to replace the original face, thus \textbf{\textit{removing }}the original identity; \textit{2)} Perturbation-based methods add quasi-imperceptible noise to disturb the judgment of  FR systems, thus \textbf{\textit{concealing}} the original identity.

Perturbation-based methods differ from synthesis-based methods in one major characteristic—\textit{they allow human  observers to see the original identity, while synthesis-based methods block it.} This characteristic helps perturbation-based methods particularly well-suited for protecting face privacy of  \textit{\textbf{subjects}} (distinct from bystanders \cite{hasan2020automatically})  in  photos, which still enables recognition by familiar friends with naked eyes, thereby preserving the photo utility without compromising its social function.   A variety of perturbation-based methods \cite{deb2020advfaces,Yang_2021_ICCV,cherepanova2021lowkey} were developed to assist individuals in protecting subject faces, e.g.,  \texttt{Fawkes} \cite{255262} (pixel level) and \texttt{ATM-GAN} \cite{Hu_2022_CVPR} (semantic level). \texttt{Fawkes} has released public software versions for  practical applications, while most methods are currently limited to the research prototype stage.

\begin{figure*}[!t]
	\centering
	\includegraphics[width=\linewidth]{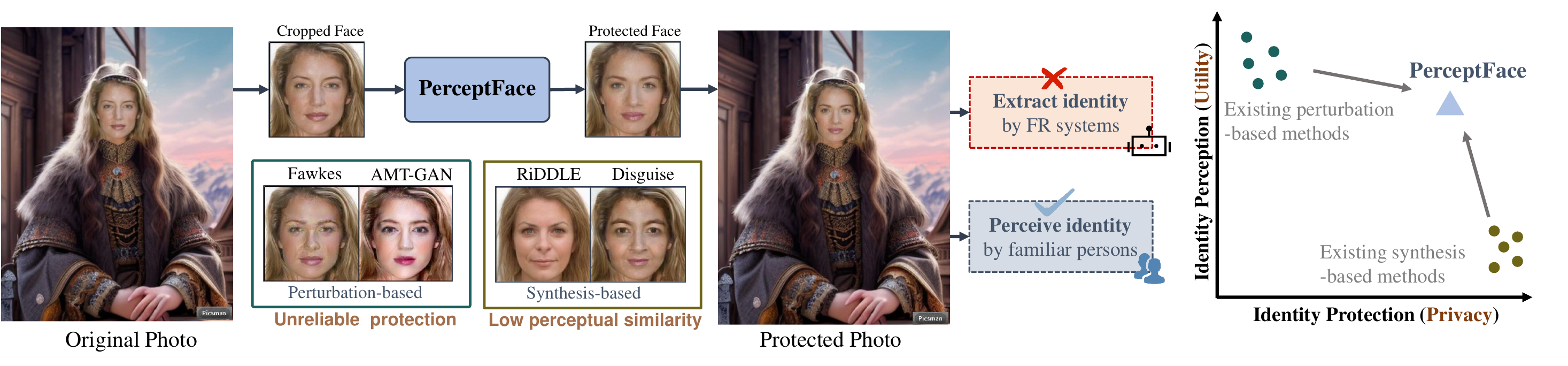}
	\caption{Illustration of the proposed PerceptFace. PerceptFace protects the subject face in the photo, making the identity unextractable yet perceptible, where the original photo is  AI-enlarged by Picsman, avoiding potential ethical and copyright issue. PerceptFace achieves a superior trade-off between privacy and utility compared to existing perturbation-based (unreliable protection) and synthesis-based methods (low perceptual similarity). }
	\label{shouyetu}
\end{figure*}



\textbf{Motivation.} It is imperative to sound the alarm for the research community that—\textit{perturbation-based methods provide just a false sense of privacy.} 
One compelling reason is the inherent unsustainability:  they defend against privacy breaches by attacking FR systems to discover vulnerabilities\cite{han2023interpreting},  but these vulnerabilities would be fixed by ever-evolving FR systems, which makes the protection just temporary. It indicates that  \textit{an attack is just an attack and not the best defense.} Thus, the research community should find new alternative solutions to protect subject face,  which was also highlighted in the recent systematization of knowledge (SoK) work\cite{wenger2023sok}.


\textbf{Our Work.} In this paper,   we explore  a promising solution (synthesis-based methods)  toward the development of more practical anti-FR tools for subject faces. Due to the change to a new identity, synthesis-based methods cannot be obstructed by the ever-evolution of FR systems. However, since facial identity and appearance are strongly correlated,  replacing a new identity would inherently result in a new facial appearance, which blocks human vision from recognizing subjects.
For this, we attempt to solve the challenge of synthesis-based methods in protecting subject face privacy, i.e., enabling familiar persons to recognize subjects.

Concretely, the major contributions of our work are summarized as follows:

\begin{itemize}
		\item{ \textit{Important Alert.} We reveal that perturbation-based methods provide just  \textit{a false sense of privacy}. In theoretical analysis, we point out four aspects, including inherent unsustainability, unattainable transferability,  weak robustness, and wrong priority. In experimental validation, we use common noise and commercial APIs to verify that the protection of mainstream methods is unreliable. Meanwhile, we alert the research community about the cautious utilization of adversarial perturbation for protecting privacy.}

        \item {\textit{Key Insight.} We present a key insight based on existing research \cite{jiang2021explainable,johnston2009familiar} and life experience, which is more aligned with the real-world scenario—\textit{In most photo sharing scenarios, the recognition of subjects relies on identity perception rather than meticulous face analysis by familiar persons. }  This insight emphasizes  ``the photo containing faces" from a global perspective,  breaking the limitation of existing methods to consider ``faces in the photo" from a local perspective.  We further divide the process of identity perception into \textit{contextual perception of non-facial regions} and \textit{coarse-grained perception of facial regions}.}


        \item {\textit{Promising  Method.} 	Based on the above insight, we propose  \textbf{\textit{the first synthesis-based method dedicated to subject faces}}, i.e., \texttt{PerceptFace}, which can allow familiar persons to perceive the subject identity while preventing  FR systems from extracting it. As shown in Fig. \ref{shouyetu}, PerceptFace achieves a superior trade-off between privacy and utility compared to existing perturbation-based (\texttt{Fawkes} \cite{255262} and  \texttt{AMT-GAN} \cite{Hu_2022_CVPR}) subject to the \textit{unreliable protection},  and synthesis-based methods (\texttt{RiDDLE } \cite{li2023riddle} and  \texttt{Disguise} \cite{cai2024disguise})  subject to the \textit{low perceptual similarity}.   According to the process of identity perception, PerceptFace benefits from \textit{attribute-preserved identity manipulation}  and   \textit{perception-enhanced identity transformation}.
        
        }

        \item {
\textit{Technical  Innovation.} We design an innovative perceptual similarity loss for faces, which promotes PerceptFace to achieve better identity perception than existing synthesis-based methods \cite{li2023riddle, cai2024disguise}.  Existing methods apply the perceptual similarity loss (mainly LPIPS) just targeted at \textit{generic categories} of images, which limits their ability to preserve identity perception. This is because human vision systems appear to allocate specialized neural resources to face perception \cite{sinha2006face}. More appropriately, our perceptual similarity loss is targeted at \textit{face categories} of images,  which can mimic the face perception of human vision by introducing perceptual sensitivity. }
        
\end{itemize}

The remainder of the paper is organized as follows.
Section \ref{Related_Work} briefly introduces the related work. Section \ref{sec3} presents the threat model and design goals. 
Section  \ref{section perturbation}  reveals the limitations of perturbation-based methods, and Section \ref{Key Thought} shows our thoughts and insights. 
Section \ref{my_method} details the proposed method.
Experimental results are presented in Section \ref{sec4} and followed
by a conclusion in Section \ref{sec5}.

\section{Related Work} \label{Related_Work}
\subsubsection{Encryption or Transformation-based Methods} Encryption-based methods utilize authorized keys to encrypt and later restore images. Tajik \textit{et al.} \cite{tajik2019balancing} developed a thumbnail-preserving image encryption scheme for cloud storage, allowing users to preview photos.   ProShare \cite{yuan2015privacy}  enables different regions of a photo to be encrypted and accessed based on user permissions. It securely encrypts images and embeds them directly within the JPEG file itself, providing users with more granular control over their online photo privacy.

In contrast, transformation-based methods do not support reversibility but offer higher efficiency.  FaceObfuscator \cite{jin2024faceobfuscator} removes key identity-related signals in the frequency domain and adopts a carefully designed obfuscation mechanism to resist reconstruction attacks. It can quickly process facial images and remove any visual information except for identity information. CamPro \cite{zhu2023campro} anonymizes facial regions at the hardware level by adjusting camera imaging parameters, providing protection against hijacking attacks. As the face has been erased during the imaging process, it provides strong privacy guarantees.

However, the protected results produced by these methods appear visually unnatural. This distinct visual discrepancy makes it easy for attackers to notice and distinguish between protected and unprotected faces. Consequently, the research community tends to favor the following two main categories of methods.

\subsubsection{ Perturbation-based Methods} Early perturbation-based methods \cite{evtimov2020foggysight,chandrasekaran2020face,li2023unganable} primarily relied on pixel-level noise to concealing identity. Fawkes \cite{255262} is a representative work in this category, offering multiple levels of noise intensity to meet diverse user privacy requirements. TIP-IM \cite{Yang_2021_ICCV} introduces a maximum mean square error criterion to suppress unnecessary noise, improving visual quality. OPOM \cite{9778974} generates a personalized face mask for each user, achieving a practical trade-off between protection effectiveness and computational efficiency. P3-Mask \cite{chow2025personalized}  builds on OPOM by integrating a focal diversity-optimized ensemble learning approach into the mask generation process, which significantly strengthens its robustness against unknown FR models.

Recent works have shifted towards semantic-level perturbations \cite{NEURIPS2022_dccbeb7a,sun2024diffam}, which provide better transferability and robustness. AMT-GAN \cite{Hu_2022_CVPR}  perturbs facial appearance via makeup synthesis to evade recognition. 3DAM-GAN \cite{lyu20233d}  leverages facial symmetry to render realistic and robust makeup, thereby enhancing the transferability to black-box models. However, despite offering improved protection performance, the method introduces noticeable visual distortions in the nasal region. ImU \cite{An.ImU.SP.2023} constructs style perturbations in the latent space of StyleGAN, while GIFT \cite{10.1145/3664647.3681344} operates in a better latent space of StyleGAN to generate global feature perturbations. DiffPrivate \cite{le2025diffprivate} further advances this direction by constructing global feature perturbations in the latent space of diffusion models (DMs), improving generalization under varied recognition settings.  To enhance security, DivTrackee \cite{fan2025divtrackee} advocates for explicitly promoting the diversity of protection results. This is achieved by building upon a text-guided image generation framework and a diversity loss based on the first-in, first-out (FIFO) principle.


\subsubsection{ Synthesis-based Methods}
Early synthesis-based methods \cite{wang2024key,chen2021perceptual,li2021identity} typically replace the entire facial region to achieve anonymization. DeepPrivacy \cite{hukkelaas2019deepprivacy} removes the original face and employs the generative adversarial network to synthesize a new face. RiDDLE \cite{li2023riddle} introduces a key-controlled replacement mechanism, enabling both reversibility and diversity in face substitution. Lopez \textit{et al. }\cite{lopez2024privacy} introduced a hardware-level method for face de-identification. Their approach first involves training an optical encoder to produce a privacy-preserving face heatmap, which is then used to generate a novel face based on a reference image.

To better preserve facial attributes, recent methods \cite{wang2024make,wen2023divide} focus on semantic-level manipulation of identity-specific features. CIAGAN \cite{maximov2020ciagan}  utilizes a conditional GAN to take identity labels as a conditional input, which allows it to generate controllable and diverse protection results. Additionally, it preserves the facial structure to ensure the protected faces remain detectable. FaceRSA \cite{zhang2024facersa} is a facial identity encryption framework that possesses all the properties of RSA. It achieves facial identity anonymization and secure de-anonymization within the decoupled StyleGAN latent space using a carefully designed cryptographic mapper. Disguise \cite{cai2024disguise}  transforms identity via differential privacy in a disentanglement-based framework, preventing re-identification attacks. AIDPro \cite{10884889} uses authentication information as a condition to guide identity transformation, enabling robust image authentication. It achieves robustness to a wide range of image distortions, such as JPEG compression and screen shooting, without needing a noise layer.

\section{Preliminary}\label{sec3}

\subsection{ System Model and Threat Model}\label{General_s}

\begin{figure}[!h]
	\centering
	\includegraphics[width=3.5in, keepaspectratio]{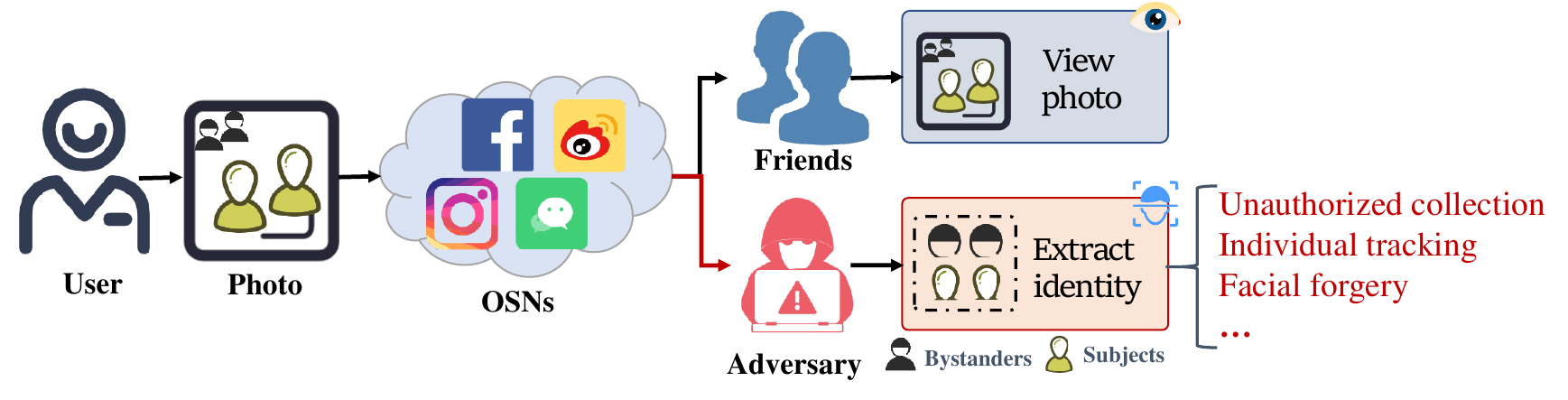}\\
	\caption{General  photo sharing scenario and threat model.}
	\label{systemmodel}
\end{figure}

We consider a general photo sharing scenario, in which a user shares a photo containing faces on online social networks (OSNs) and then friends can view the photo to feel the conveyed message. During the sharing process, photos may undergo compression, re-encoding, or other image processing operations by the social platform.

Our threat model focuses on \textbf{identity extraction by FR tools}, which can be used for malicious purposes, e.g., unauthorized collection, individual tracking, or facial forgery, as shown in Fig. \ref{systemmodel}. Given the widespread availability of high-performance pre-trained FR tools (advanced models or commercial APIs), adversaries can readily access them to facilitate such a malicious goal.

 In our work, identity inference by professional attackers is not considered. Although some non-FR tools can infer the subject's identity based on contextual cues such as location or clothing, they are incapable of obtaining facial information. Moreover, lacking direct visual evidence, such identity inference has low accuracy and credibility. 





\subsection{Face Category and Anti-FR Methods}

We classify the faces to be protected in photos, as they have different protection requirements.

\begin{itemize}[leftmargin=*,itemsep=0pt, parsep=0pt]
	\item{\textit{Subject Face:} Subjects are the main persons in a photo. The subjects often occupy a prominent position in the photo and maybe the user themselves, a close friend, or a family member.   Subject faces are not allowed to be significantly modified,   because friends  \underline{care} who the subject is.  }
	\item{\textit{Bystander Face:} Bystanders  \cite{hasan2020automatically} are other people who appear unintentionally in a photo. They are not the subject of the photo and maybe casual passers-by, other participants in an event, or strangers in a public place. Bystander faces are allowed to be significantly modified because friends \underline{do not care} who the bystander is.}	
\end{itemize}

Setting aside methods  that degrade visual quality \cite{yang2022study,pias2022decaying,10816466}, existing anti-FR methods  can be mainly  categorized:   

\begin{itemize}[leftmargin=*,itemsep=0pt, parsep=0pt]
	\item{\textit{Perturbation-based Method:} Such methods add quasi-imperceptible noise to disturb the judgment of  FR models, thus concealing the original identity.  The noise  		
		can be pixel level (e.g., \texttt{Fawkes} \cite{255262} )  or  semantic level (e.g., \texttt{ATM-GAN} \cite{Hu_2022_CVPR})  .  Since this noise has a weak effect on the facial appearance, human vision \underline{can observe} the original identity. }
	
	\item{\textit{Synthesis-based Method:} Such methods generate a face with a new identity to replace the original face, thus removing the original identity. The generated face can be entirely new (e.g., \texttt{RiDDLE} \cite{li2023riddle} )  or only altered in identity (e.g., \texttt{Disguise} \cite{cai2024disguise}). Since a new identitity brings a new facial appearance, human vision \underline{cannot observe} the original identity.}	
\end{itemize}

\begin{figure}[!h]
	\centering
	\includegraphics[width=3in, keepaspectratio]{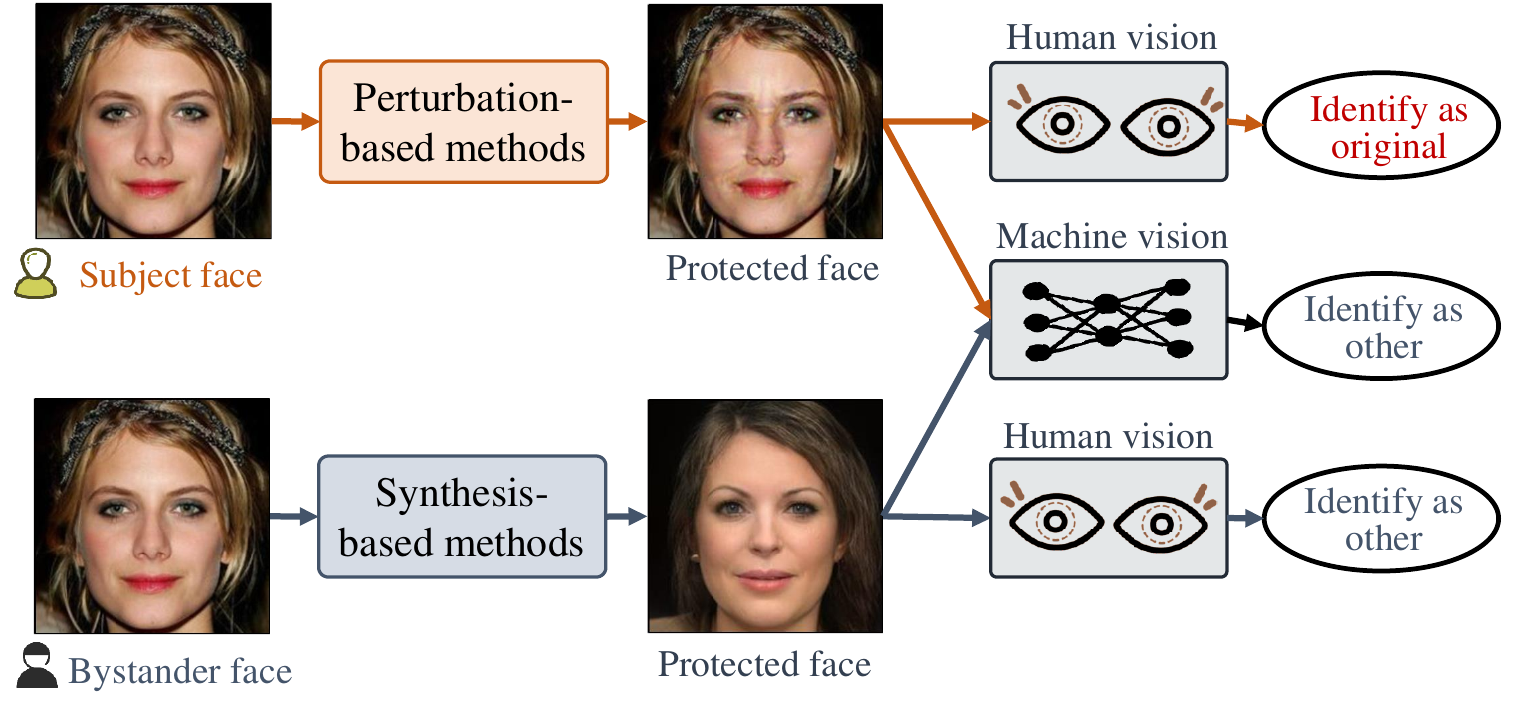}\\
	\caption{Corresponding anti-FR Methods applicable to different categories of faces. Perturbation-based methods are applied to subject faces, and synthesis-based methods are applied to bystander faces. Distinctly, our method is the \textit{first} synthesis-based method dedicated to subject faces.}
	\label{comparing_effect}
\end{figure}

Based on the above, perturbation-based methods are applicable to protect subject face privacy, and synthesis-based methods are applicable to protect bystander face privacy, which are visually displayed in Fig. \ref{comparing_effect}. Distinctively, our method is \textit{the first synthesis-based method}, which is designed for subject face privacy rather than bystander face.

\subsection{Design Goals}
Our work is to design an anti-FR tool to protect  \textit{subject faces }in photos, as we reveal that perturbation-based methods cannot reliably protect them in Section \ref{section perturbation}. {Without destroying  non-cropped areas in the photo}, the design goals are as follows:

\begin{tcolorbox}
	[breakable,		                    
	arc=0mm, auto outer arc,            
	boxrule= 0pt,                        
	boxsep = 0mm,                       
	left = 1mm, right = 1mm, top = 1mm, bottom = 1mm, 
	]
	{
		\textbf{Goal} \ding{182}:\textbf{ {Preserving the utility that familiar persons \textbf{can recognize} the subject via human vision.}}  Anti-FR tools should not compromise the social utility of photos. Although these tools modify only the \textit{cropped facial region}, such alterations may hinder friends from visually recognizing the subject, thereby impairing the intended usage of the photo in social contexts. Thus, it is essential to allow recognition by familiar persons.
		
		
		
	}
\end{tcolorbox}	

\begin{tcolorbox}
	[breakable,		                    
	arc=0mm, auto outer arc,            
	boxrule= 0pt,                        
	boxsep = 0mm,                       
	left = 1mm, right = 1mm, top = 1mm, bottom = 1mm, 
	]
	{
		\textbf{Goal} \ding{183}:\textbf{{ Protecting the privacy that FR systems \textbf{ cannot extract identity} of the subject via machine vision.} }
		Facial identity is highly sensitive, and its leakage can pose severe privacy risks. With the high accuracy and wide accessibility of deep learning-based FR tools, facial identity is increasingly vulnerable to unauthorized extraction.  Thus, it is essential to prevent identity extraction by FR systems.
	}
\end{tcolorbox}

\subsection{Dodging or Impersonation Protection}

According to whether the protected identity is specific, existing protection methods can be categorized into \textit{dodging protection} and \textit{impersonation protection}. The former
 enables the protected identity to be a  \textit{non-targeted identity}, while the latter enables it to be a \textit{targeted identity}.
	
\textit{	Our work focuses on dodging protection} and we argues that impersonation protection is less suitable in practice:
	
\subsubsection{Risk of Re-Identification} Prior research \cite{zhou2024rethinking} has shown that high impersonation success   may coincide with low dodging success. This implies that even if the protected identity is close to the target identity, it may also remain  highly  similar to the original identity. As a result, adversaries may still re-identify the original individual by selecting the top-2 most similar identities. Therefore, impersonation protection may pose a significant privacy risk of re-identification.
	
\subsubsection{Misuse of Target Identity} Impersonation protection requires specifying a target identity, which may correspond to a real individual for malicious purposes, raising concerns over portrait rights and ethical implications.  For instance, a user or an insider might deliberately set a celebrity as the target identity and fabricate  evidence implicating them in illegal activities. Even if such allegations are later refuted, the incident may cause reputational harm and public distress to the celebrity. Therefore, impersonation protection may raise ethical  concerns regarding   identity misuse.

\section{Perturbation-based Methods Cannot Reliably Protect Subject Face Privacy} \label{section perturbation}
\subsection{Intuitive Effectiveness}

\begin{figure}[!h]
	\centering
	\includegraphics[width=\linewidth]{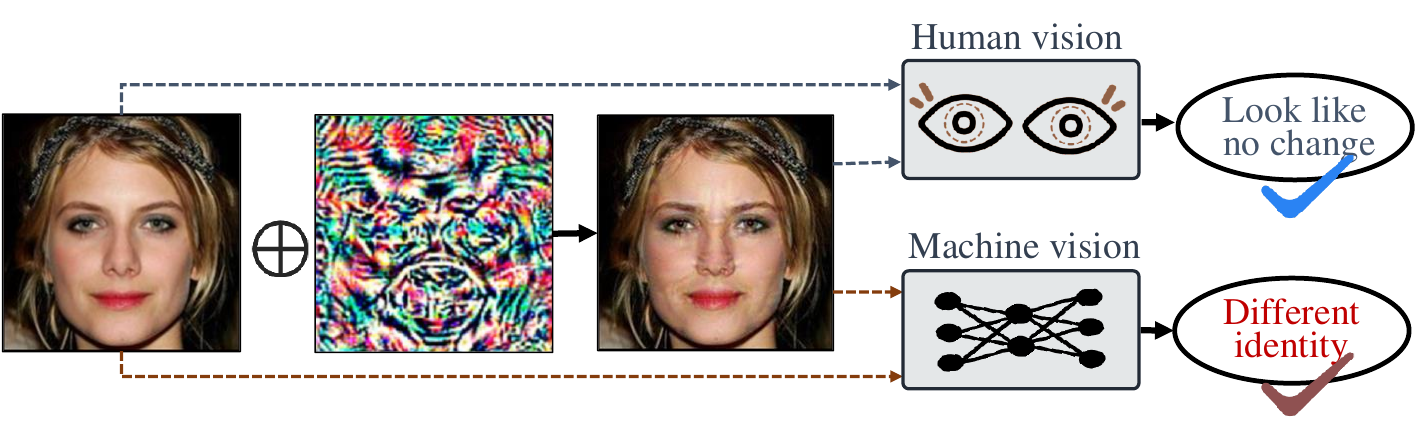}
	\caption{Intuitive effect of perturbation-based methods.}
	\label{ap_visual}
\end{figure}

Adversarial perturbations are crafted quasi-imperceptible noises that are added to the input data, causing deep learning models to make incorrect prediction. Because of the slight impact on human visual perception, adversarial perturbations can be applied in subject face privacy protection, called \textit{perturbation-based methods. }

Given a face image $x$, the crafted adversarial perturbation $\Delta x$ can prevent the automatic identity extraction from the unauthorized FR model $f(\cdot)$. Perturbation-based methods can be formalized to solve a constrained optimization problem:
\begin{equation}\label{eq1}
	\begin{aligned}&
		\max_{\Delta x}\mathcal{D}(f(x+\Delta x),f(x)), \\& ~~~~~\mathrm{~s.t.~}\parallel \Delta x\parallel_p\leq \epsilon,
	\end{aligned}
\end{equation}
where $\mathcal{D}(\cdot, \cdot)$ is a metric of identity distance (typically Euclidean or cosine distance),  $\parallel \cdot \parallel_p$ denotes the $l_p$ norm  (typically $p=1,2,\infty$), and $\epsilon$ constrains the maximum extent of the perturbation deviation. In Equation \ref{eq1}, maximizing the objective function can achieve	\textbf{Goal} \ding{183}, and the constraint term can promise	\textbf{Goal} \ding{182}.

Through solving the above optimization problem, $\Delta x$ that satisfies the requirements can be constructed to protect face privacy. As shown in Fig. \ref{ap_visual}, the protected image is only added with a small level of noise, making human vision think that the image looks the same before and after the protection. Therefore,  familiar persons can still recognize the subject identity (\textbf{Goal} \ding{182}). In addition, machine vision is disturbed by perturbations and recognizes the face in the protected image as others. Therefore, FR tools cannot extract facial identity,  which achieves  (\textbf{Goal} \ding{183}).

\subsection{A False Sense of Privacy}
Thanks to the almost unchanged visual content, perturbation-based methods can easily maintain identity consistency for human vision. However, we point out that it just provides a false sense of privacy,  which cannot stand up to real-world challenges.

\begin{enumerate}
\item \textbf{Inherent Unsustainability:} Perturbation-based methods rely on exploiting the vulnerabilities of deep neural networks to deceive FR systems \cite{han2023interpreting}. However, these vulnerabilities are not static; modern FR systems are continuously updated and retrained to fix known weaknesses. As a result, adversarial perturbation that once succeeded may lose their effectiveness as the systems evolve. This leads to a persistent \textit{cat-and-mouse game}—as new perturbations are created, new defenses emerge, rendering past efforts obsolete. Thus, such methods fail to provide sustainable protection for users.

\item \textbf{ Unattainable Transferability:}  It is unattainable to craft a universal adversarial perturbation that can counteract all FR systems. Different FR systems have differences in architectural design, training strategy, data distribution, or optimization.  These factors result in different decision boundaries and vulnerabilities. While some studies attempt to improve transferability \cite{Li_2023_CVPR}, their success is often limited to specific model families.  
Thus, such methods fail to provide transferable protection protection for users.
\item \textbf{ Weak Robustness:} It is difficult for adversarial perturbations to survive in the OSN transmission. OSNs perform various image processing on photos shared by users to improve loading speed, save bandwidth, adapt to device screens, etc. Common operations like JPEG compression would alter the image pixels, which may remove the adversarial perturbations.  While the employment of noise layers \cite{shin2017jpeg} improves robustness to noise, it is impossible to enumerate all the noises. 
Thus, such methods fail to provide robust protection for users.

\item 	
\textbf{Wrong Priority:} 	As defined in Equation \ref{eq1}, the constraint is designed for utility,  which gives utility a higher priority than privacy. That is, privacy is not necessarily satisfied in the optimized result.  We argue that this prioritization is flawed: the risk of facial identity leakage typically outweighs the minor loss in visual fidelity.  Therefore, the optimization problem Equation \ref{eq1} should be reset to privacy as the constraint and utility as the optimization objective.  Thus, such methods fail to provide prior protection for users.
\end{enumerate}

\subsection{ Evaluation of Reliable Protection }

In the previous subsection, we analyzed the limitations of perturbation-based methods in protection from four aspects. This section verifies the above conclusion through experiments. Specifically, we select three representative methods for evaluation, including:

\begin{itemize}[leftmargin=*,itemsep=0pt, parsep=0pt]
	\item{\texttt{Fawkes \cite{255262}:}   Fawkes is a popular perturbation-based protection system, which can generate quasi-imperceptible noise for individual images.  The team of Fawkes also provides applications for Windows and MacOS, which drive the progress of privacy protection for everyone. Fakwes contains three levels of protection to trade off privacy and utility. To adapt to real-world application scenarios, we use the middle level of protection for  evaluation.}

	\item{\texttt{OPOM \cite{9778974}:} Unlike image-specific Fawkes, OPOM constructs user-specific perturbations by exploiting the identity space of a single user, which can protect all the face images of that user. In addition, it uses momentum enhancement methods and DFANet to improve the transferability. Based on the recommendation in the paper, we set $\epsilon$ to 8 to carry out the evaluation. } 
	
	\item{\texttt{AMT-GAN \cite{Hu_2022_CVPR}:} Unlike Fawkes and OPOM which both use \textit{pixel-level noise }as perturbation, AMT-GAN employs  \textit{semantic-level noise} (makeup) to enhance visual naturalness. In addition, it employs an ensemble training strategy with input diversity enhancement to enhance the transferability and robustness. We carried out the evaluation based on the pre-trained model provided by the authors.
	} 
\end{itemize}

There are also some related methods that were not selected for testing because of similar design ideas or non-open-source codes, while the chosen methods were sufficiently representative of them.  

We conducted two types of evaluations, including adding noise (for robustness) and utilizing commercial APIs (for transferability and sustainability). Please note that priority is intuitive and thus doesn't require to be evaluated

\begin{figure}[!t]
	\centering
	\includegraphics[width=\linewidth]{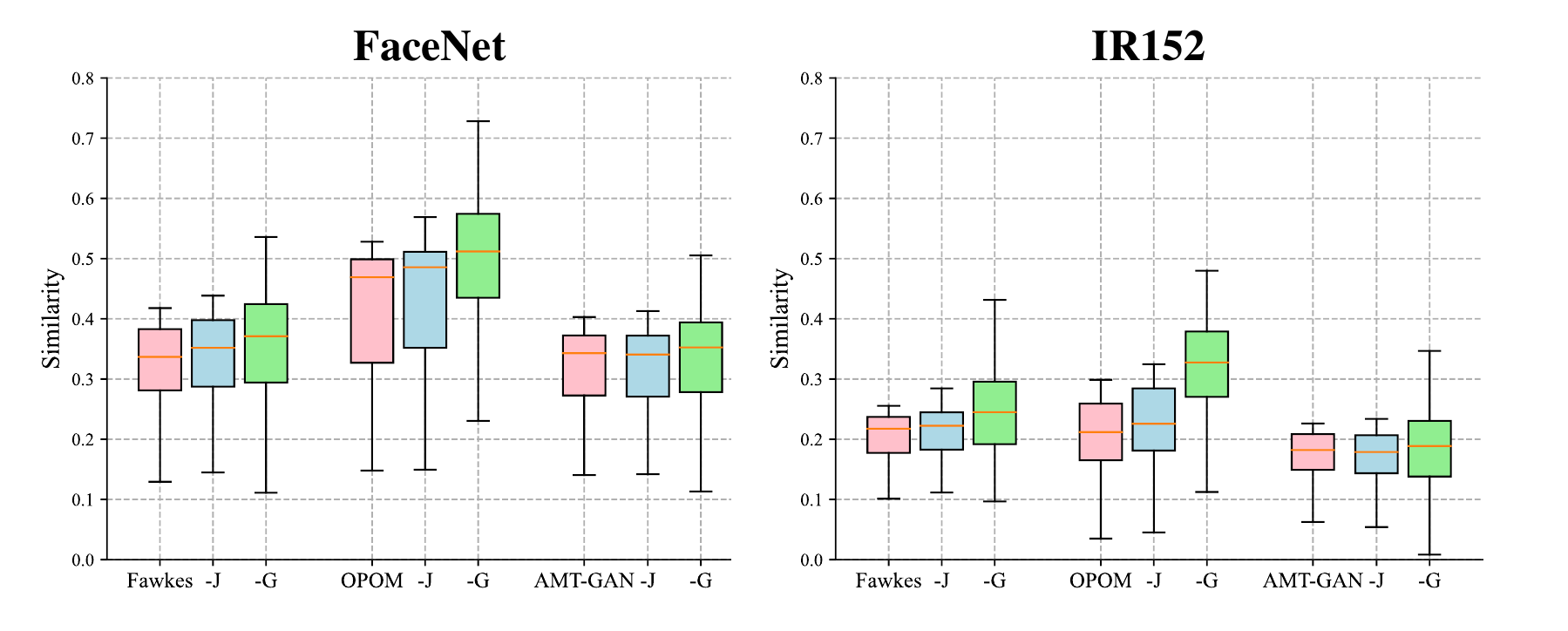}
	\caption{Perturbation-based against noise, where "-J" means JPEG compression and "-G" means Gaussian noise.}
	\label{AP_robust}
\end{figure}

\textbf{ Against Noise.}  To reduce the bandwidth requirements for storage and transmission, uploaded photos are typically JPEG compressed. Meanwhile, during transmissions, random noise may be introduced, which often exhibits Gaussian distribution. To evaluate the \textit{robustness}, we apply JPEG compression ($Q=90$) and  Gaussian noise ($\sigma=0.1$).

Specifically, we selected FaceNet \cite{schroff2015facenet} and IR152 \cite{he2016deep} to evaluate identity change before and after protection. To demonstrate the effect of noise, we selected 100 test samples where the perturbation had a large change in identity. Fig. \ref{AP_robust}  shows box-and-line plots of cosine similarity. Fawkes and OPOM are obviously affected by JPEG compression and Gaussian noise, resulting in a degradation of protection performance. This is because the perturbations generated by them are inherently pixel-level noise, which is very easily weakened or corrupted by pixel changes caused by JPEG compression and Gaussian noise.  AMT-GAN utilizes makeup as a natural perturbation, which avoids the effect of JPEG compression but still brings a weak effect from Gaussian noise.  Therefore, the reliable protection of perturbation-based methods is insufficient when faced with noise.

\begin{figure}[!t]
	\centering
	\includegraphics[width=\linewidth]{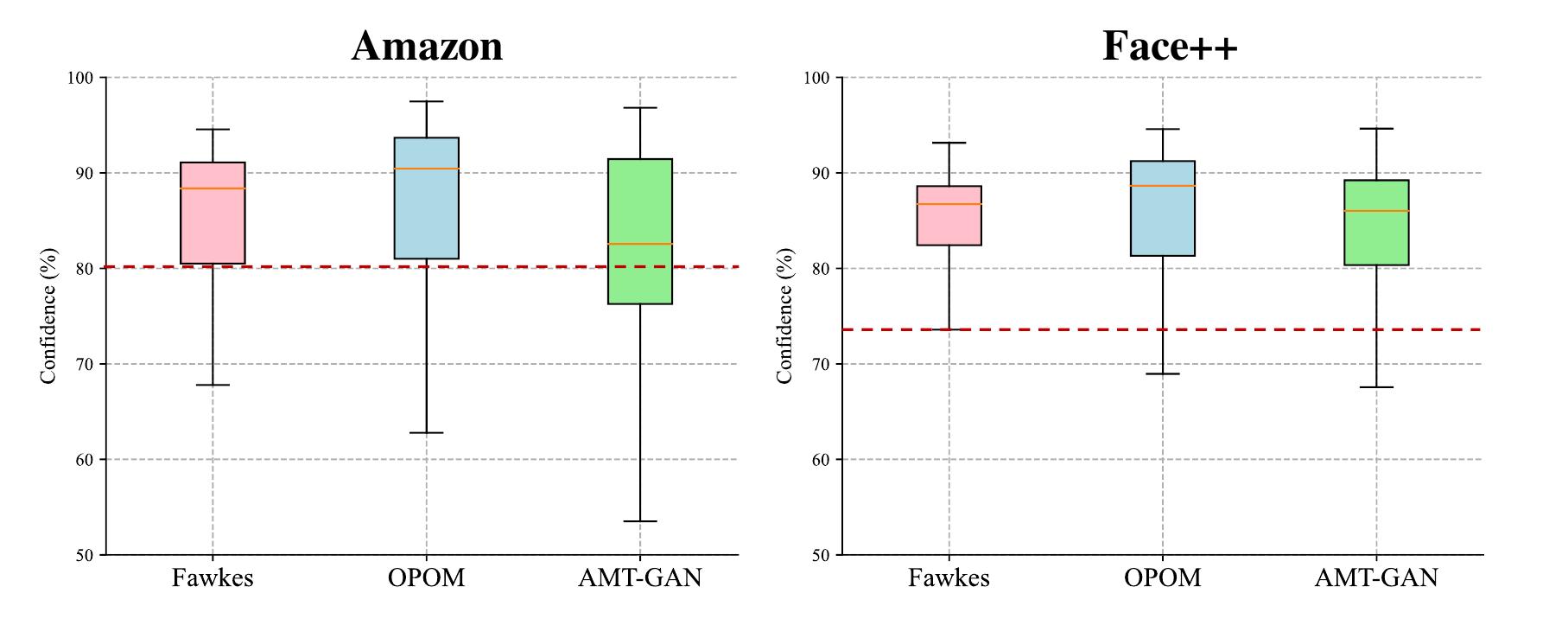}
	\caption{Perturbation-based methods against commercial APIs. Most of the results are above the thresholds (red line).}
	\label{AP_API}
\end{figure}

\textbf{Against Commercial APIs.} Advanced FR APIs do not disclose their models and parameters, making it difficult for existing techniques to consider them as surrogate models for perturbation generation. Therefore, these APIs can be well used to evaluate the \textit{transferability} of privacy protection in real scenarios.  In particular, commercial APIs periodically patch existing vulnerabilities to improve identification performance, which can help us evaluate the \textit{sustainability }of privacy protection. 

Specifically, we selected  Amazon and Face++ to evaluate identity change before and after protection, where the matching threshold of Amazon is 80.00\% and Face++ is 74.00\%. As shown in Fig. \ref{AP_API}, we can find that OPOM  has the worst identity protection. Fawkes will also update its systems to improve transferability, but brings only a weak effect. AMT-GAN further improves the transferability through  integration strategies, which provides the best protection. However, most of the results are above the thresholds, allowing commercial APIs to extract real identity features as well. Although these methods were effective in resisting commercial APIs at the time in the original papers, their protection performance has significantly declined with updates and patches to the commercial  APIs, rendering them only temporarily effective.

\begin{figure}[!t]
	\centering
	\includegraphics[width=\linewidth]{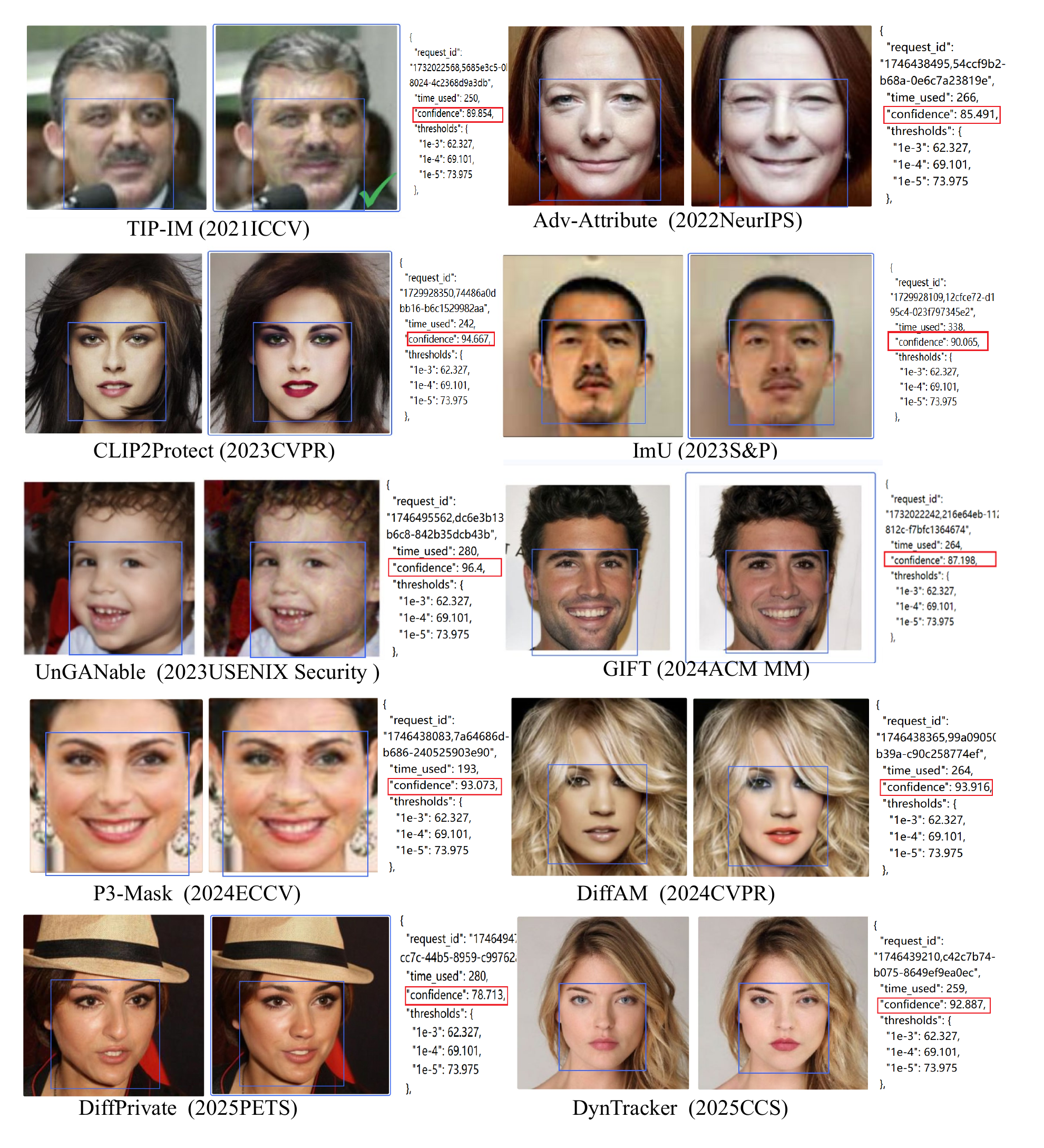}
	\caption{Other perturbation-based methods against Face++. All of the results were above the threshold (74\%) and thus privacy is not protected.}
	\label{AP_API_other}
\end{figure}

In addition, we cropped images from other similar methods for testing, e.g., TIP-IM \cite{Yang_2021_ICCV}, ImU \cite{An.ImU.SP.2023}, CLIP2Protect \cite{shamshad2023clip2protect},  and GIFT \cite{10.1145/3664647.3681344}. In Fig. \ref{AP_API_other}, the results show that these methods still keep a high confidence of face matching, thus enabling to extract personal identity.   Therefore, due to periodic vulnerability fixes, the protection is no longer reliable against the ever-evolving commercial APIs.

\section{ Key Thought and  Insight} \label{Key Thought}

\subsection{Is It Feasible for Synthesis-based Methods to Protect Subject Face Privacy?}

Previous section reveals that adversarial perturbation cannot provide reliable privacy protection. It is worth exploring whether synthesis-based methods can be an alternative solution to protect subject face privacy.

\textbf{Advantage.} Synthesis-based methods can achieve reliable privacy protection for \textbf{Goal }\ding{183}. 

\begin{itemize}[leftmargin=*,itemsep=0pt, parsep=0pt]
\item \textit{ Sustainability}: They do not depend on vulnerabilities in FR systems, so identity protection still works even if the vulnerabilities are fixed.

\item  \textit{ Transferability:} The goal of all FR systems is to extract the identity in a face, so they are all capable of extracting the changed identity. 

\item \textit{ Robustness:} As a high-level semantic \cite{he2016deep,10578062}, facial identity can show excellent robustness, so the extraction of the changed identity will not be significantly affected by common noises. 

\item \textit{Priority:} They often minimize alterations to identity-unrelated attributes while changing identity,
i.e., prioritizing privacy over utility. 

\item \textit{Imperceptibility:}  Compared to the irregular noise introduced by perturbation-based methods, synthesis-based methods provide distributions that are indistinguishable from the real data, which can be more imperceptible to the adversary.

\item \textit{Understandability:} Non-expert users can understand their privacy-preserving mechanism, i.e., changing identity, whereas users cannot understand how noise or make-up can attack FR systems.
\end{itemize}

\textbf{Challenge.} However, synthesis-based methods cannot achieve  satisfactory utility for \textbf{Goal} \ding{182}. The main reason is that facial appearance is closely correlated with identity. The change in identity must modify facial appearance (e.g., skin color or eye shape), thus preventing familiar persons from recognizing subjects.  It should be noted that all the existing synthesis-based methods \cite{maximov2020ciagan,kuang2024facial,barattin2023attribute} aim to anonymize visual appearance to block human vision.  Although some works \cite{li2023riddle,lopez2024privacy}  employ LPIPS to enhance visual similarity, the effect they bring is weak. Therefore, existing synthesis-based methods cannot be directly applied to subject face privacy protection, and it is challenging to achieve \textbf{Goal} \ding{182}.

\textbf{Thought.} Firstly, almost all works ignore a crucial point—Not ``faces in the photo" but  ``the photo containing faces". Existing methods deal with cropped faces locally, but neglect to consider the global impact on the photo after protection.  In real scenarios, the main task of users is feeling the global message conveyed by the photo rather than the identity represented by the face. Therefore, a weak alteration of faces may not matter.

Secondly,  the recognition by familiar persons is robust to image alterations and better than the recognition by unfamiliar persons \cite{chapman2018robust,kramer2018understanding}. Perturbation-based methods bring subtle noise, which we consider to be \textit{stringent} in real scenarios.  In fact, even if the nose, eyes, and mouth of the face are altered \textit{more relative to the noise}, familiar persons still can recognize the subject by viewing the photo.	


Therefore, synthesis-based methods are\textit{\textbf{ feasible to protect subject face privacy}}, as long as the alteration of facial appearance is not sufficient to affect the recognition by human vision. For this, we should analyze how familiar persons recognize subjects in photos.

\subsection{How Familiar Persons Recognize Subjects in  Photo Sharing Scenario?}

 

 Based on  existing research and life experience, we present the following insight:

\begin{tcolorbox}
	[breakable,		                    
	colback= blue!10!white,		            
	arc=0mm, auto outer arc,            
	boxrule= 0pt,                        
	boxsep = 0mm,                       
	left = 1mm, right = 1mm, top = 1mm, bottom = 1mm, 
	]
	{\textbf{Insight:} \textit{ In most photo sharing scenarios, the recognition of subjects by familiar persons relies on \textbf{identity perception} rather than \textbf{meticulous face analysis}.}}
\end{tcolorbox}

\textbf{Supporting Evidence.} We support the above insight from three perspectives.

\begin{itemize}[leftmargin=*,itemsep=0pt, parsep=0pt]

\item  \textit{ From the perspective of photo  content, }  most shared photos don't emphasize faces as the primary visual content (except for self-portraits) \cite{hu2014we}.  Instead, these photos typically highlight activities and events \cite{khosla2014makes}, e.g., playing sports, attending gatherings, or documenting daily routines, where the face may be small, partially visible, or even occluded. Therefore, the main task of friends is to interpret the social  activity or event depicted in the photo, rather than to engage in face analysis, naturally facilitates identity perception by familiar persons.

\item \textit{From the perspective of cognitive, } familiar persons build stable mental representations of others through repeated social exposure. These representations integrate various personal characteristics \cite{zhang2015beyond} beyond facial geometry — such as clothing style, body posture, and even makeup. As such, familiar viewers can still reliably recognize subjects in degraded or partial-view conditions \cite{burton1999face}. In addition,  recognition by familiar persons is more robust to image degradation than that by unfamiliar persons \cite{johnston2009familiar,ramon2018familiarity}.
This also indicates that familiar persons do not rely on meticulous face analysis; otherwise, their recognition performance would be comparable to that of unfamiliar persons.

\item \textit{From the perspective of social context,} photo sharing often occurs within well-defined interpersonal boundaries—such as among friends or family—where the identity of the sharer and expected subjects are largely known. In these cases, recognition is more about confirming the presence of expected individuals than identifying unknown persons. Contextual factors \cite{backstrom2007wherefore, oh2016faceless} like accompanying text, shared event themes, or the photo’s time of posting provide additional verification cues. The act of recognition in such contexts is often anticipatory and guided by shared background knowledge, which reduces the requirement for meticulous face analysis.

\end{itemize}

\begin{figure}[!t]
	\centering
	\includegraphics[width=0.9\linewidth]{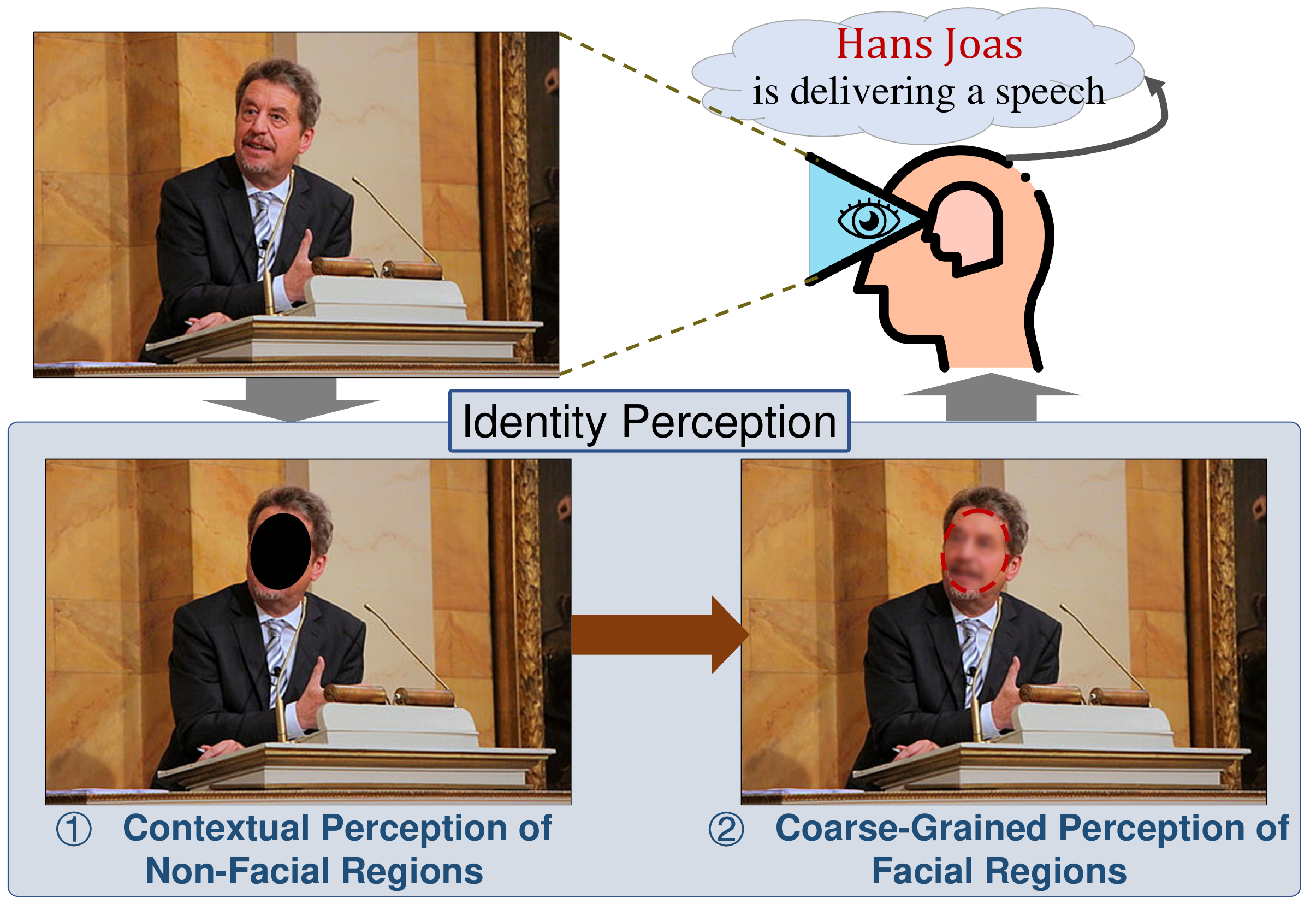}
	\caption{A sample of identity perception, where the photo of Hans Joas is from IMDB-WIKI. The face region is blurred (red dashed line) to mimic coarse-grained perception.}
	\label{identity_perception}
\end{figure}


We  define identity perception:

\begin{tcolorbox}
	[breakable,		                    
	colback= orange!10!white,		            
	arc=0mm, auto outer arc,            
	boxrule= 0pt,                        
	boxsep = 0mm,                       
	left = 1mm, right = 1mm, top = 1mm, bottom = 1mm, 
	]
	{\textbf{Definition:} \textit{\textbf{ Identity perception }  refers to the quick and intuitive cognitive process by which familiar persons infer a person’s identity  under non-strictly recognizable scenarios, combining contextual perception of non-facial regions and coarse-grained perception of facial regions, without meticulous face analysis.}}
\end{tcolorbox}

\textbf{Detail.} We detail the two sub-processes: \textit{1) Contextual perception of non-facial regions.} Apart from non-visual information (e.g., social context), appearance context (e.g.,  hairstyle, clothing, and physique) can stimulate the prior knowledge of friends and help them to quickly associate the closest identity. Especially for highly familiar subjects, even if their faces are not visible, friends can accurately recognize the subject through these appearance context. \textit{2) Coarse-grained perception of facial regions.} Coarse-grained facial characteristics (e.g., general contours, facial structure, and skin color) further help friends to convince the subject identity. In real-world scenarios, most friends just coarsely view the facial region without carefully comparing the facial details (e.g., precise nose size). 
Fig. \ref{identity_perception} presents a sample process with \textit{Hans Joas}.


\begin{figure*}[!t]
	\centering
	\includegraphics[width=\linewidth]{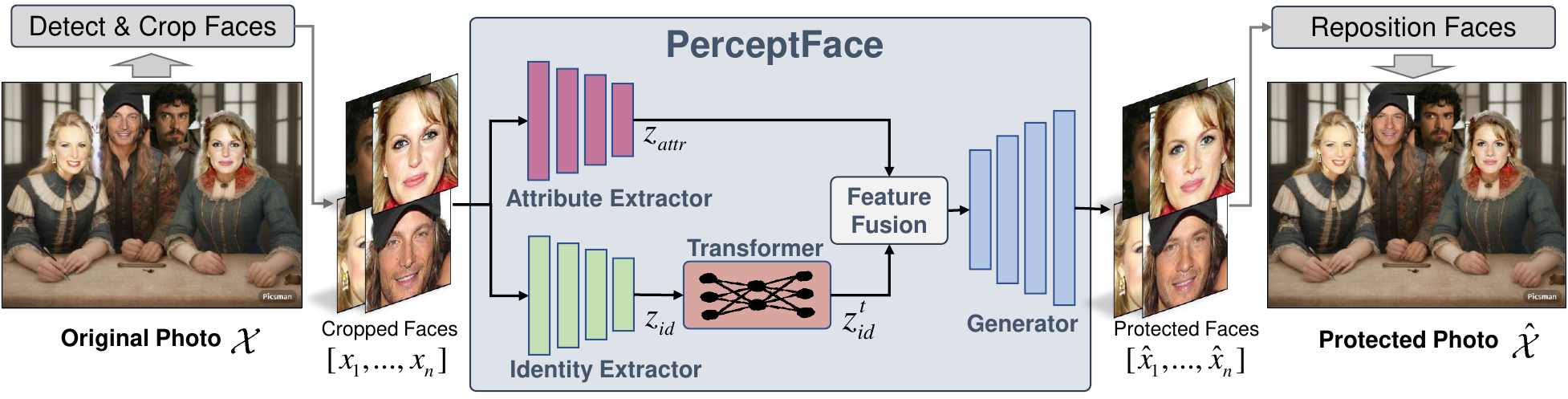}
	\caption{The pipeline of PerceptFace, where the original photo is  AI-enlarged by Picsman, avoiding potential ethical and copyright issue. Firstly, the subject faces are detected and cropped. Secondly, PerceptFace protects each subject face. Finally, the protected faces are repositioned to the original photo. Since PerceptFace only slightly modifies the facial area for \textit{identity protection}, which allows familiar persons to \textit{perceive identity} of the subject. }
	\label{fangantu}
\end{figure*}

\section{Methodology} \label{my_method}

\subsection{Pipeline of PerceptFace}
Based on the \textit{insight} in the previous section, we make a modification to the design \textbf{Goal} \ding{182}. Then, our goals are:

\begin{tcolorbox}
	[breakable,		                    
	colback= red!10!white,		            
	arc=0mm, auto outer arc,            
	boxrule= 0pt,                        
	boxsep = 0mm,                       
	left = 1mm, right = 1mm, top = 1mm, bottom = 1mm, 
	]
	{\textbf{Goal} \ding{182}*: \textit{ Preserving the utility that familiar persons  \textbf{can perceive identity} of the subject via human vision. }
		
		\textbf{Goal} \ding{183}\;: \textit{ Protecting the privacy that FR systems  \textbf{ cannot  extract  identity} of the subject via machine vision. }
	}
\end{tcolorbox}

To achieve the above goals, we propose  the first synthesis-based method for subject face privacy \textit{i.e.}, PerceptFace. Given a photo $\mathcal{X}$, the  PerceptFace can edit the subject faces $[ x_1, ..., x_n ]$ in it, enabling the corresponding identity to be changed while maintaining a high level of visual perceptual similarity. When sharing the protected photo $\mathcal{\hat{X}}$, the sharer's friends can easily perceive the subject identity via human vision (\textbf{Goal} \ding{182}*), but the FR systems cannot extract accurately the identity via machine vision (\textbf{Goal} \ding{183}).

Fig. \ref{fangantu} shows the usage flow of the proposed PerceptFace in privacy-preserving social photo sharing.

\textbf{1) Firstly}, we detect and crop the subject faces in the photo.  Since distinguishing between the subjects and bystanders is not the focus of our work, for the sake of simplicity, we assume that all persons in the photo are subjects. 
\begin{equation}
	[ x_1, ..., x_n]=FaceDetect(\mathcal{X}),
\end{equation}
where $FaceDetect(\cdot)$ is the face detector.


\textbf{2) Secondly}, for each subject face, the proposed PerceptFace is utilized to protect them. By elaborately changing the identity features, the protected face is only slightly different from the original face in visual perception, while it belongs to another identity in machine vision,
\begin{equation}
	\hat{x}_i = PerceptFace(x_i), i=1,..., n.
\end{equation}

\textbf{3) Finally}, all of the protected faces are repositioned into the original photo, obtaining the protected photo $\mathcal{\hat{X}}$. Since the background areas of the protected faces remain undistorted, they are able to blend naturally into the original photo,

\begin{equation}\mathcal{\hat{X}}
	=FaceReposit(\mathcal{X}, [\hat{x}_1, ..., \hat{x}_n]).
\end{equation}

For the protected photo, since the contextual information is fully preserved and the facial region is only slightly altered, the familiar persons can perceive the subject identity with their prior knowledge, thus preserving utility.  Benefiting from the robustness of high-level semantic (identity) change, arbitrary FR systems can only extract the changed (not the original) identity in noisy environments, thus protecting privacy.

\subsection{PerceptFace}

\textbf{Design Idea.} PerceptFace aims to make identity unextractable yet perceptible.  As a synthesis-based method, PerceptFace is able to make identity not extractable, so the challenge lies in making identity perceptible. Following the process of identity perception mentioned in the previous section, our idea is to design:

\begin{itemize}[leftmargin=*,itemsep=0pt]
	\item{\textit{	For contextual perception of non-facial regions}, we disentangle identity and attribute features, which can reduce the changes of non-facial regions when protecting identity. \newline (\textit{ Attribute-Preserved  Identity Manipulation})}
	\item{\textit{For coarse-grained perception of facial regions}, we design an innovative perceptual similarity loss for faces, which can maintain high similarity perception of facial regions. \newline (\textit{Perception-Enhanced Identity Transformation})} 
\end{itemize}



\textbf{Flow.} As shown in Fig. \ref{fangantu}, PerceptFace firstly extracts the attribute features $z_{attr}$ and identity features $z_{id}$ of the cropped face $x$ by the attribute extractor $E_{attr}$ and identity extractor $E_{id}$, respectively. Secondly, $z_{id}$ is elaborately modified via the designed transformer $T$ to obtain the protected version $z^t_{id}$. Thirdly, $z^t_{id}$ is fused with the original attribute $z_{attr}$,  and then synthesized into the protected face $\hat{x}$ by the generator.  In this way, the protected face maintains  a similar visual appearance with the original one, while the identity information is effectively protected. Note that\textit{ due to information loss during synthesis, the identity features  $\hat{z}_{id}$  of $\hat{x}$  typically exhibit slight deviations from $z^t_{id}$.}

\textbf{Network Structure.} The \textit{attribute extractor} $E_{attr}$ adopts four convolutional layers for downsampling, followed by a batch normalization layer and a ReLU activation layer. We use the off-the-shelf pre-trained Arcface as the \textit{identity encoder} $E_{id}$, which can accurately extract identity features. To fuse identity and attribute features, we introduce the ID Injection module \cite{chen2020simswap} that can facilitate the generator to synthesize faces with the target identity and attributes. Similar to $E_{attr}$, the generator $G$ adopts four deconvolutional layers for upsampling, followed by a batch normalization layer and a ReLU activation layer. The input and output of the \textit{transformer} $T$ are one-dimensional vectors of the same length, thus a simple two-layer perceptron can be its backbone. Lastly, $L_2$ regularization is followed by the last layer to constrain identity features in hyperspherical spaces.

\subsection{Attribute-Preserved  Identity Manipulation}

Attribute-preserved identity manipulation (APIM) aims to preserve attributes while manipulating identity. Feature disentanglement can separate the face features into attribute features and identity features, which facilitates the preservation of appearance perception when manipulating identity. By changing only the identity features, face privacy can be protected while identity-irrelevant attributes are preserved, thus maintaining contextual information, e.g., background and hairstyle. 

Since the identity extractor is pre-trained, it already has the ability to extract identity features accurately. Therefore, the remaining work is to train both the attribute extractor $E_{attr}$ and the generator $G$ to disentangle the attribute features and synthesize the specified face.  Similar to the existing work \cite{ cai2024disguise,chen2020simswap}, we use the following losses:

\textbf{Identity Loss.} Identity loss forces the identity $\hat{z}_{id}$ of the synthetic face to be consistent with the transformed identity $z^t_{id}$. Cosine similarity is used to measure identity loss:
\begin{equation}
	\mathcal{L}_{id}=1-\frac{\hat{z}_{id}\cdot z^t_{id}}{\|\hat{z}_{id}\|_2\|z^t_{id}\|_2}.
\end{equation}

\textbf{Attribute Loss.} Attribute loss forces the attribute of the synthetic face to be consistent with the original face.  Weak-feature matching loss \cite{chen2020simswap} is used to measure attribute loss:
\begin{equation}
	\mathcal{L}_{attr}=\sum_{i=h}^H\frac1{N_i}||D_i(x)-D_i(\hat{x}))||_1,
\end{equation}
where $D_i(\cdot)$ represents the feature extracted from the $i$-th layer by  the discriminator $D$, $N_i$ indicates the number of  elements in the $i$-th layer,  $H$ denotes the total number of layers, and $h$ specifies the starting layer for calculating the attribute loss.

\textbf{Fusion Loss.} The fusion loss can guide the attributes and modified identity to synthesize into the specified face.  Since the specified face is not known in advance, the original face is usually used as the specified face. In other words, the original face is disentangled and then reconstructed back to the original face, which uses the unmodified identity features and attribute features. Specifically, the $L_1$ loss is used to construct a pixel-level fusion loss.
\begin{equation}
	\mathcal{L}_{fus}=\|G(z_{id}, z_{attr})-x\|_1.
\end{equation}

\textbf{Adversarial loss.}  To improve visual quality, adversarial loss $\mathcal{L}_{adv}$  \cite{NIPS2017_892c3b1c} is used to make synthetic faces indistinguishable from real faces.

\textbf{Objective \uppercase\expandafter{\romannumeral1}. } Overall, the training  objective of APIM  is
formulated as follows:
\begin{equation} \mathcal{L}_{total}^{\uppercase\expandafter{\romannumeral1}}= \mathcal{L}_{adv}+\lambda_{id}  \mathcal{L}_{id}+ \lambda_{attr}  \mathcal{L}_{attr}+\lambda_{fus}\mathcal{L}_{fus},
\end{equation}
where $\lambda_{*}$ are hyperparameters which control the magnitude of different losses.

\begin{figure}[!b]
	\centering
	\includegraphics[width=\linewidth]{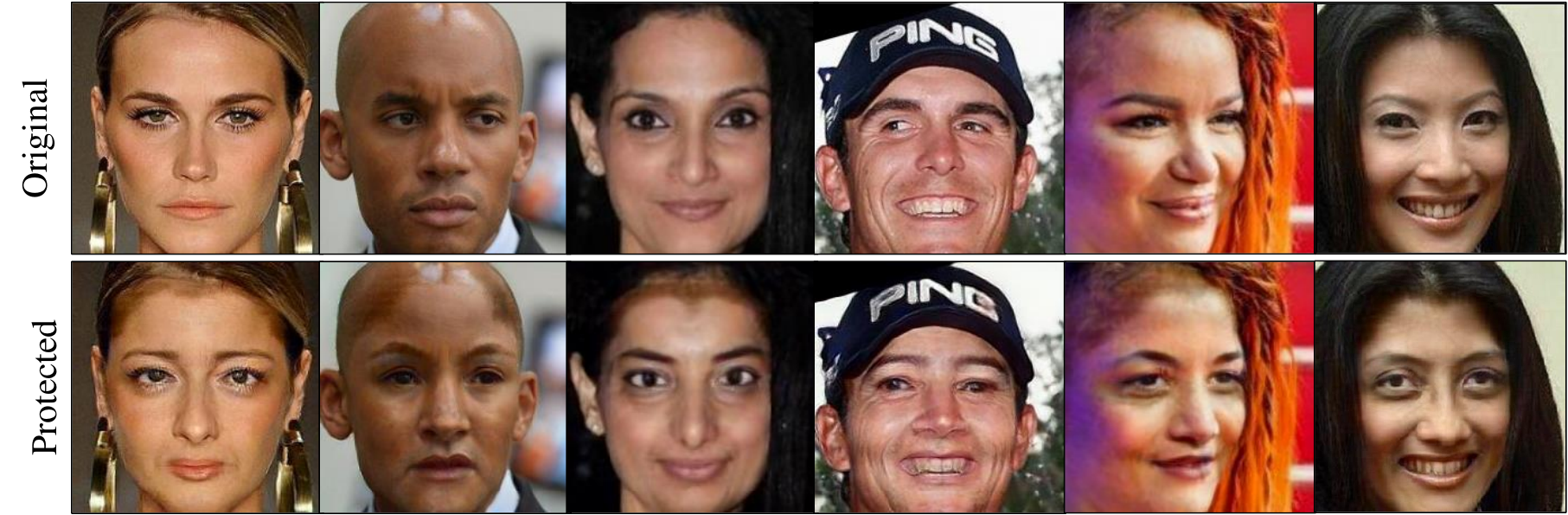}
	\caption{Protected results via APIM. Although the non-facial areas are preserved, the visual changes in the facial areas are still easily perceived. }
	\label{IT_1}
\end{figure}

\subsection{Perception-Enhanced Identity Transformation} APIM in the previous subsection can preserve identity-irrelevant attributes (mainly non-facial regions), but this is not sufficient for maintaining identity perception consistency. As identity is likewise correlated with facial regions, arbitrary changes in identity can lead to distortions in the facial area, which can bring about obvious differences in visual perception, e.g., general contours, structure, and skin color. As shown in Fig. \ref{IT_1}, it can be observed that the visual appearance of the protected face is significantly altered, which is adequate to affect the identity perception by familiar persons.



In this section, perception-enhanced identity transformation (PEIT) works on constructing  an identity transformation to delicately edit the identity features, so as to maintain high visual perceptual similarity while changing the identity, which can be formalized as,
\begin{equation}\label{eq2}
	\begin{aligned}
		&\max_{z^t_{id}}\mathcal{S}(G(z^t_{id},z_{attr}),x), \\ &~~\mathrm{~s.t.~}    cos(\hat{z}_{id}, z_{id}) <\tau,
	\end{aligned}	
\end{equation}
where $\mathcal{S}(\cdot, \cdot)$ is a metric of perceptual similarity,  $cos(\cdot, \cdot)$ is the cosine similarity, and $\tau$ is the threshold. Compared to Eq. \ref{eq1},  Eq. \ref{eq2} treats privacy as a constraint to maximize the objective function (utility), which is more in line with the goal of privacy protection.

\begin{tcolorbox}
	[breakable,		                    
	arc=0mm, auto outer arc,            
	boxrule= 0pt,                        
	boxsep = 0mm,                       
	left = 1mm, right = 1mm, top = 1mm, bottom = 1mm, 
	]
	{
		\textbf{Feasibility Analysis}: Unless doing a facial analysis task, human vision usually does not pay more attention to fine-grained facial features. In the real world, we also often find examples of similar identities in human visual perception, especially between twins or between celebrities and specific public. Nevertheless, upon meticulous facial analysis, it can be distinguished that these perceptually similar identities are different, e.g., different sizes of noses or eye distances.  Therefore, it is feasible to maintain high visual perceptual similarity while changing the identity.
	}
\end{tcolorbox}

To avoid the time-consuming problem caused by the iterative solution for single data, we train an MLP-based transformer $T$ to provide real-time identity transformation, which  is supervised by the following losses:

\textbf{Enhanced Face Perception for Utility.} To allow identity perception, the protected face should maintain a high visual perceptual similarity to the original face. However, it is difficult to directly measure visual face perception, which is a complex neural inference process.

As an alternative approach, learned perceptual image patch similarity (LPIPS) \cite{zhang2018unreasonable} is a perceptual similarity metric for generalized categories of images.  Unlike the simple or low-level metrics (e.g., PSNR and SSIM), LPIPS utilizes the variability of depth features to measure perceptual similarity, which is more in alignment with human vision in texture, color, and structure. Therefore, we consider using LPIPS as the perceptual similarity loss.
\begin{equation}
	\mathcal{L}_{lpips}=\|\mathcal{P}(x)-\mathcal{P}(\hat{x})\|_2,
\end{equation}
where $\mathcal{P}(\cdot)$ is the pre-trained perceptual feature extractor, which is selected as AlexNet in our work. However, LPIPS targets generic category images, rather than face category images. Previous research \cite{sinha2006face} has shown that the human vision system appears to allocate specialized neural resources to face perception. Therefore, the ability of LPIPS to mimic then face perception is limited.  


Existing research \cite{abudarham2016reverse}  pointed out that human vision has different perceptual sensitivity (PS) to different facial features via user study.  Inspired by it, we design a new perceptual similarity loss for faces, which introduces PS to face regions based on LPIPS.  Regions with higher PS would limit smaller changes.  

\begin{figure}[!t]
	\centering
	\includegraphics[width=0.6\linewidth]{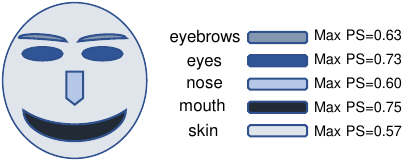}
	\caption{  Perceptual sensitivity (PS) of different facial regions, where  darker colors indicate greater PS.}
	\label{facePS}
\end{figure}

Specifically, a pre-trained face parser (FP) is adopted to segment out face regions $[M_1, ..., M_k ]$, representing the masks of eyebrows, eyes, nose, mouth, and skin,
\begin{equation}
	[M_1, ..., M_k ]=FP(x). 
\end{equation}
It should be noted that non-facial areas (e.g., hair) are not included, as these are preserved to a greater extent through APIM.  Then, we calculate the corresponding region changes,
\begin{equation}
	\mathcal{L}_{region}=\sum _{i=1}^{k}  \alpha_i \|  M_i \odot x-M_i \odot \hat{x}  \|_2,
\end{equation}
where the value of $\alpha_i$  is taken as the maximum  PS of the different features in the region in \cite{abudarham2016reverse}, as shown in Fig.\ref{facePS}. Specially, 
\begin{equation}
	\alpha_i = \frac{\max_{f \in \mathcal{F}_i} \text{PS}(f)}{\sum_j \max_{f \in \mathcal{F}_j} \text{PS}(f)},
\end{equation}
where $\mathcal{F}_i$ denotes the set of facial features in region $M_i$, and $\text{PS}(f)$ is the PS  score of feature $f$ in $M_i$, as reported in \cite{abudarham2016reverse}. For example, the eye region includes features such as eye color (PS = 0.73) and eye shape (PS = 0.61), and we choose the maximum value (0.73) as the representative PS for that region. Finally, the obtained values are normalized to produce the final perceptual sensitivity weights: $\alpha = [0.192, 0.223, 0.183, 0.229, 0.174]$.

Eventually, the designed  perceptual similarity  loss for faces is defined as,
\begin{equation}
	\mathcal{L}_{per}=\mathcal{L}_{lpips}+\mathcal{L}_{region}.
\end{equation}

\textbf{Basic Identity Deviation for Privacy.} To ensure identity protection, the faces to be protected should be moved away from the original face in the identity feature space, effectively minimizing their similarity. Specifically, when the similarity drops below a predefined threshold $\epsilon$, no further reduction in loss is necessary. Therefore, the identity deviation loss can be expressed as:
\begin{equation}
	\mathcal{L}_{pri}=max(\epsilon, cos(E_{id}(x). \; E_{id}(\hat{x}))).
\end{equation}
Only a single identity extractor is employed, which is sufficient to change identity. Of course, integrating results from multiple identity extractors is also recommended to avoid model bias, but not necessary.

\textbf{Objective \uppercase\expandafter{\romannumeral2}. } Overall, the training  objective of PEIT  is
formulated as follows:
\begin{equation}  \mathcal{L}_{total}^{\uppercase\expandafter{\romannumeral2}}= \lambda_{pri}  \mathcal{L}_{pri}+ \lambda_{per}  \mathcal{L}_{per},
\end{equation}
where $\lambda_{*}$ are hyperparameters which control the magnitude of different losses.

\section{Experimental Results}\label{sec4}
\subsection{Setup}

\subsubsection{Datasets} \texttt{VGGFace2} is a large-scale face dataset comprising over 3.31 million images of 9,131 identities, featuring significant variations in pose, age, illumination, ethnicity, and profession. All images are resized to $224 \times 224$. We use the face images of 8,631 identities for training, while the remaining identities are reserved for testing. \texttt{CelebA-HQ}, widely utilized for face recognition tasks, includes 30,000 aligned face images at a resolution of $1024 \times 1024$. Each image is annotated with 5 landmarks and 40 binary attributes. Similarly, we resize these images to $224 \times 224$. To assess the generalizability of our method to other datasets, CelebA-HQ is exclusively used for testing. 

Since the social photo sharing scenario involves sharing photos that contain faces (rather than cropped faces), we use uncropped photos (IMDB-WIKI) for the rest of the experiment.

\subsubsection{Dual-phase Training Strategy}
To avoid conflicts caused by multi-objective optimization, we adopt a two-stage training strategy to learn the parameters of PerceptFace. \textit{1) In the first stage}, we utilize 	\textit{Objective \uppercase\expandafter{\romannumeral1} } to  train $E_{attr}$ and $G$ to achieve attribute-preserved identity manipulation. Specifically,  both $E_{attr}$ and $G$ are optimized with Adam optimizer with $(\beta_1=0.5, \beta_2=0.99)$,  the initial learning rate is set to 0.0004, and 
the hyperparameters are set to $\lambda_{id}=30$, $\lambda_{attr}=10$, and $\lambda_{fus}=10$. \textit{2) In the second stage}, we utilize  \textit{Objective \uppercase\expandafter{\romannumeral2} } train $T$  to achieve perception-enhanced identity transformation.  Specifically,  $T$ is optimized with Adam optimizer with $(\beta_1=0.99, \beta_2=0.99)$,  the initial learning rate is set to 0.0004, and 
the hyperparameters are set to $\lambda_{pri}=5$ and  $\lambda_{per}=5$.

\subsubsection{Baselines}  We selected  representative synthesis-based methods, i.e.,  RiDDLE  \cite{li2023riddle} (entirely new face) and  Disguise \cite{cai2024disguise} (only altered in identity). \texttt{RiDDLE} is a de-identification model that synthesizes the entire head, which also synthesizes non-facial regions (e.g., hair) to provide enhanced visual anonymity. It also can restore the original face through the correct password. \texttt{Disguise} employs feature disentanglement to independently change identity to preserve more attributes, which is more consistent with our work. In addition, we also selected the   representative perturbation-based methods, \texttt{Fawkes} \cite{255262} (pixel-level) and  \texttt{AMT-GAN} \cite{Hu_2022_CVPR} (semantic-level),  to show the protection capability of our method.

\begin{figure}[!t]
	\centering
	\includegraphics[width=\linewidth]{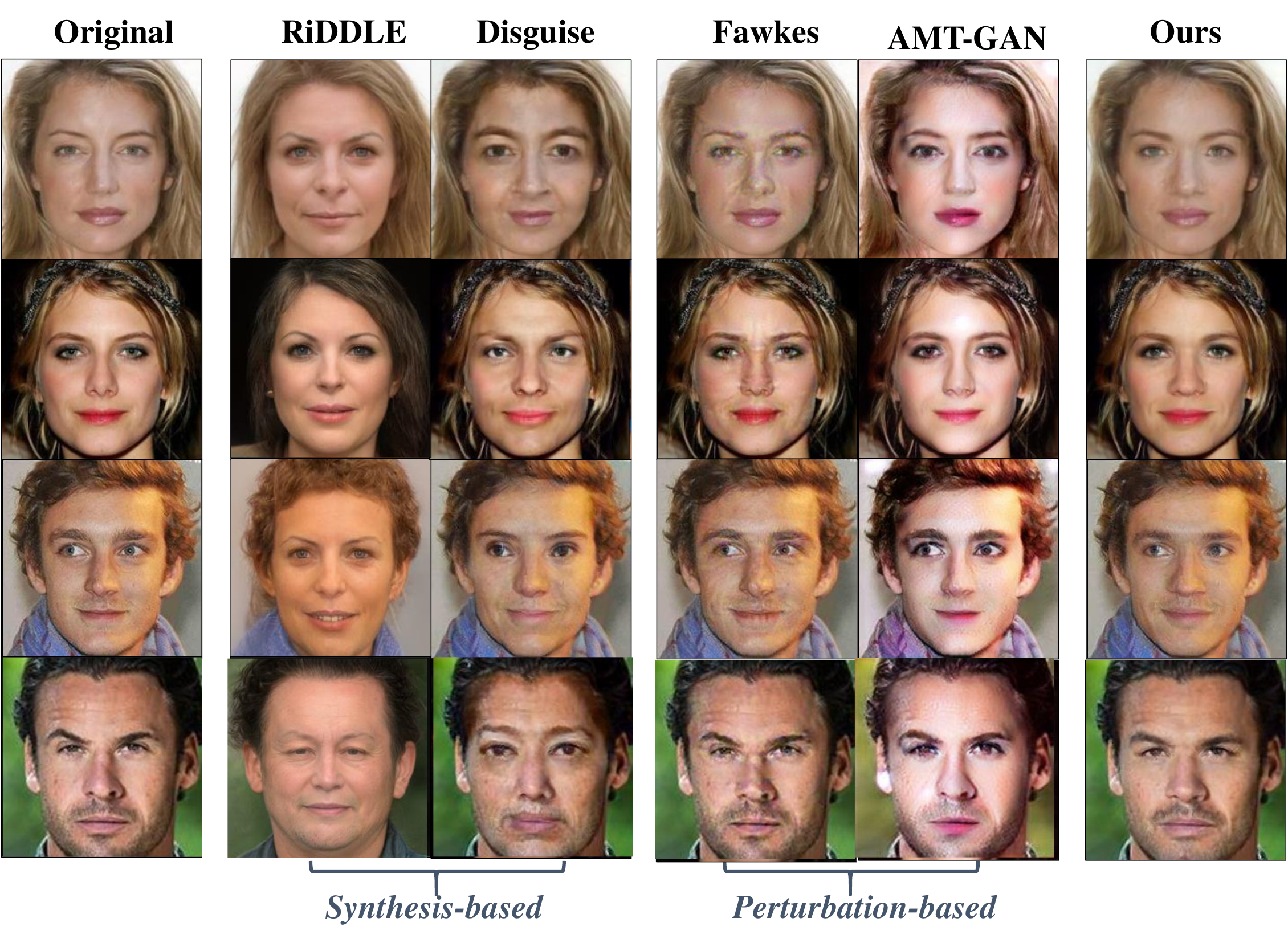}
	\caption{Visual results of our method and baselines. Compared to synthesis-based methods, ours has higher perceptual similarity. Compared to perturbation-based methods, ours has higher visual naturalness.}
	\label{visual_results_}
\end{figure}

\begin{table*}[!t]
	\begin{center}
		\caption{Identity similarity before and after protection under different  FR tools.}
		
		\label{ID_similarity}
		\scalebox{1}{
			\begin{tabular}{ccccccc|ccccc}
				\toprule	
				&	&\multicolumn{5}{c|}{VGGFace2}&\multicolumn{5}{c}{CelebA-HQ}\\   
				\cmidrule(r){3-7}  \cmidrule(r){8-12}
				&&\multicolumn{2}{c}{Synthesis-based }&\multicolumn{2}{c}{Perturbation-based} 				&&\multicolumn{2}{c}{Synthesis-based }&\multicolumn{2}{c}{Perturbation-based}\\
				\cmidrule(r){3-4}  \cmidrule(r){5-6} \cmidrule(r){8-9}\cmidrule(r){10-11}
				&	&RiDDLE&Disguise&Fawkes&AMT-GAN&Ours	&RiDDLE&Disguise&Fawkes&AMT-GAN&Ours\\ 
				\midrule
				
				\multirow{4}{*}{Models}&		FaceNet&\cellcolor{darkblue!35}0.1029&\cellcolor{darkblue!50}-0.0171&0.5495&0.6746&\cellcolor{darkblue!20}0.3729&\cellcolor{darkblue!35}0.0758&\cellcolor{darkblue!50}-0.0396&0.5497&0.5367&\cellcolor{darkblue!20}0.3063\\
				
				&IR152&\cellcolor{darkblue!20}0.0291&\cellcolor{darkblue!50}-0.0692&0.4420&0.5238&\cellcolor{darkblue!35}0.0288&\cellcolor{darkblue!20}0.0915&\cellcolor{darkblue!50}-0.0841&0.4450&0.3586&\cellcolor{darkblue!35}0.0209	\\
				
				&IRSE50&\cellcolor{darkblue!35}0.0714&\cellcolor{darkblue!50}-0.0961&0.6728&0.6776&\cellcolor{darkblue!20}0.1179&\cellcolor{darkblue!20}0.1889&\cellcolor{darkblue!50}-0.0824&0.6688&0.5739&\cellcolor{darkblue!35}0.1414\\
				&	MobileFace&\cellcolor{darkblue!35}0.1292&\cellcolor{darkblue!50}-0.0396&0.6888&0.6767&\cellcolor{darkblue!20}0.2269&\cellcolor{darkblue!35}0.2199&\cellcolor{darkblue!50}0.0171&0.6835&0.6175&\cellcolor{darkblue!20}0.2838\\
				
				\hline
				\multirow{2}{*}{APIs}
				
				&	Amazon&\cellcolor{darkblue!35}0.1252&\cellcolor{darkblue!50}0.0936&0.8543&0.8540&\cellcolor{darkblue!20}0.3858&\cellcolor{darkblue!35}0.1340&\cellcolor{darkblue!50}0.0669&0.8764&0.7719&\cellcolor{darkblue!20}0.3421\\
				&	Face++&\cellcolor{darkblue!35}0.3057&\cellcolor{darkblue!50}0.2592&0.8942&0.8641&\cellcolor{darkblue!20}0.5319&\cellcolor{darkblue!35}0.4311&\cellcolor{darkblue!50}0.2861&0.8834&0.8038&\cellcolor{darkblue!20}0.5267\\

				\bottomrule
			\end{tabular}
		}
	\end{center}	
\end{table*}

\begin{table*}[!t]
	\begin{center}
		\caption{Protection success rates under different FR tools.}
		\label{psr}
		\scalebox{1}{
			\begin{tabular}{ccccccc|ccccc}
				\toprule	
				&	&\multicolumn{5}{c|}{VGGFace2}&\multicolumn{5}{c}{CelebA-HQ}\\   
				\cmidrule(r){3-7}  \cmidrule(r){8-12}
				&&\multicolumn{2}{c}{Synthesis-based }&\multicolumn{2}{c}{Perturbation-based} 				&&\multicolumn{2}{c}{Synthesis-based }&\multicolumn{2}{c}{Perturbation-based}\\
				\cmidrule(r){3-4}  \cmidrule(r){5-6} \cmidrule(r){8-9}\cmidrule(r){10-11}
				
				&	&RiDDLE&Disguise&Fawkes&AMT-GAN&Ours	&RiDDLE&Disguise&Fawkes&AMT-GAN&Ours\\ 
				\midrule

				\multirow{4}{*}{Models}&		FaceNet&\cellcolor{darkblue!35}99.40\%&\cellcolor{darkblue!50}100.00\%&58.20\%&26.20\%&\cellcolor{darkblue!20}93.70\%&\cellcolor{darkblue!35}99.90\%&\cellcolor{darkblue!50}100.0\%&54.00\%&62.20\%&\cellcolor{darkblue!20}96.60\%\\
				
				&IR152&\cellcolor{darkblue!35}99.80\%&\cellcolor{darkblue!50}99.90\%&8.00\%&3.40\%&\cellcolor{darkblue!20}98.90\%&\cellcolor{darkblue!20}92.70\%&\cellcolor{darkblue!50}99.60\%&7.60\%&19.60\%&\cellcolor{darkblue!35}98.40\%	\\

				&IRSE50&\cellcolor{darkblue!35}98.60\%&\cellcolor{darkblue!50}99.90\%&0.40\%&0.20\%&\cellcolor{darkblue!20}96.90\%&\cellcolor{darkblue!35}95.70\%&\cellcolor{darkblue!50}99.30\%&0.50\%&1.70\%&\cellcolor{darkblue!20}93.00\%\\
				&	MobileFace&\cellcolor{darkblue!35}98.70\%&\cellcolor{darkblue!50}99.70\%&0.30\%&1.00\%&\cellcolor{darkblue!20}92.60\%&\cellcolor{darkblue!35}93.20\%&\cellcolor{darkblue!50}98.80\%&0.20\%&0.50\%&\cellcolor{darkblue!20}90.70\%\\
				
				\hline
				\multirow{2}{*}{APIs}	
				
				&	Amazon&\cellcolor{darkblue!50}100.00\%&\cellcolor{darkblue!50}100.00\%&19.00\%&25.00\%&\cellcolor{darkblue!50}100.00\%&\cellcolor{darkblue!50}100.00\%&\cellcolor{darkblue!50}100.00\%&17.00\%&41.00\%&\cellcolor{darkblue!50}100.00\%\\
				
				&	Face++&\cellcolor{darkblue!50}100.00\%&\cellcolor{darkblue!50}100.00\%&2.00\%&5.00\%&\cellcolor{darkblue!35}99.00\%&\cellcolor{darkblue!50}100.00\%&\cellcolor{darkblue!50}100.00\%&3.00\%&7.00\%&\cellcolor{darkblue!35}99.50\%\\
				
				\bottomrule
				
			\end{tabular}
		}
	\end{center}	
\end{table*}

\subsection{Evaluation of  Privacy Protection}
The privacy protection goal is to
prevent the FR systems from extracting the original identity of the face. Fig. \ref{visual_results_} illustrates the protection results of our method and baselines. To evaluate privacy, we utilize  FR tools to extract the identity features of the faces before and after protection, respectively, and then compare their similarity.  Advanced FR models and APIs  are adopted as the  tools, including IR152, IRSE50, FaceNet, MobileFace, Amazon\footnote{https://aws.amazon.com/cn/rekognition/}, and Face++\footnote{https://www.faceplusplus.com.cn/face-comparing/}, where the models adopt cosine similarity and the APIs use the self-defined similarity. 

\textit{	1) PerceptFace exhibits significantly stronger privacy protection performance than perturbation-based methods, and comparable performance compared to existing synthesis-based methods.}

\begin{figure}[!t]
	\centering
	\includegraphics[width=\linewidth]{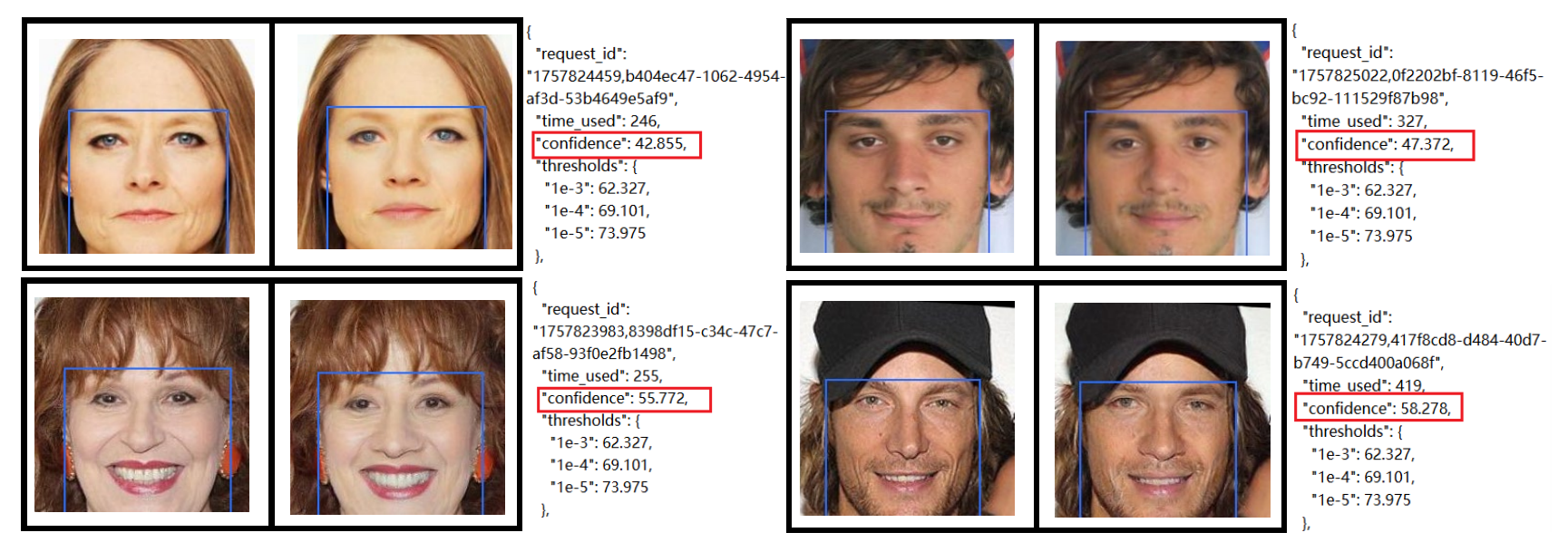}
	\caption{Some test results in Face++, presenting the  effectiveness and  practicality of our method.}
	\label{fujian_Face++}
\end{figure}

\textbf{Low Identity Similarity.} Table \ref{ID_similarity} shows the identity similarity before and after protection. Figure  \ref{fujian_Face++} shows examples of PerceptFace by Face++. The top three best performances are \textit{labeled blue}, and darker colors indicate better protection. Disguise is the most effective because of the significant modifications made to the identity features. RiDDLE, although it generates a completely new head, brings about a relatively small change in identity compared to Disguise because of the need to support reversibility. Compared to these synthetic-based methods, our method has a relatively weak identity deviation capability, but also significantly reduces identity similarity. In particular, our method also outperforms RiDDLE in some of the results. These perturbation-based methods also have the ability to reduce the identity similarity but this is too weak compared to our method.

\textbf{High PSRs.} In Table \ref{psr}, we further calculated the protection success rate (PSR) based on the matching thresholds at 0.001 FAR for all FR models. 
Consistent with the results analyzed in Table \ref{ID_similarity}, the best performance is still achieved by these synthesis-based methods. Our method also achieves a high PSR of more than 90\%, which is much higher than that of the perturbation-based method. Perturbation-based methods achieve relatively high PSRs only on specific FR tools but extremely low results on others, which also reveals that these methods have poor transferability.  Furthermore, the protection performance of our method drops a bit on IRSE50 and MobileFace. This is due to the fact that their model architectures are smaller, making them also focus on coarse-grained face perception to some extent. While ever-evolving FR tools will use more refined networks and thus focus on fine-grained facial analysis, e.g., Face++.  Therefore, the high accuracy of our method on APIs indicates the effectiveness of its protection.


\begin{figure}[!t]
	\centering
	\includegraphics[width=\linewidth]{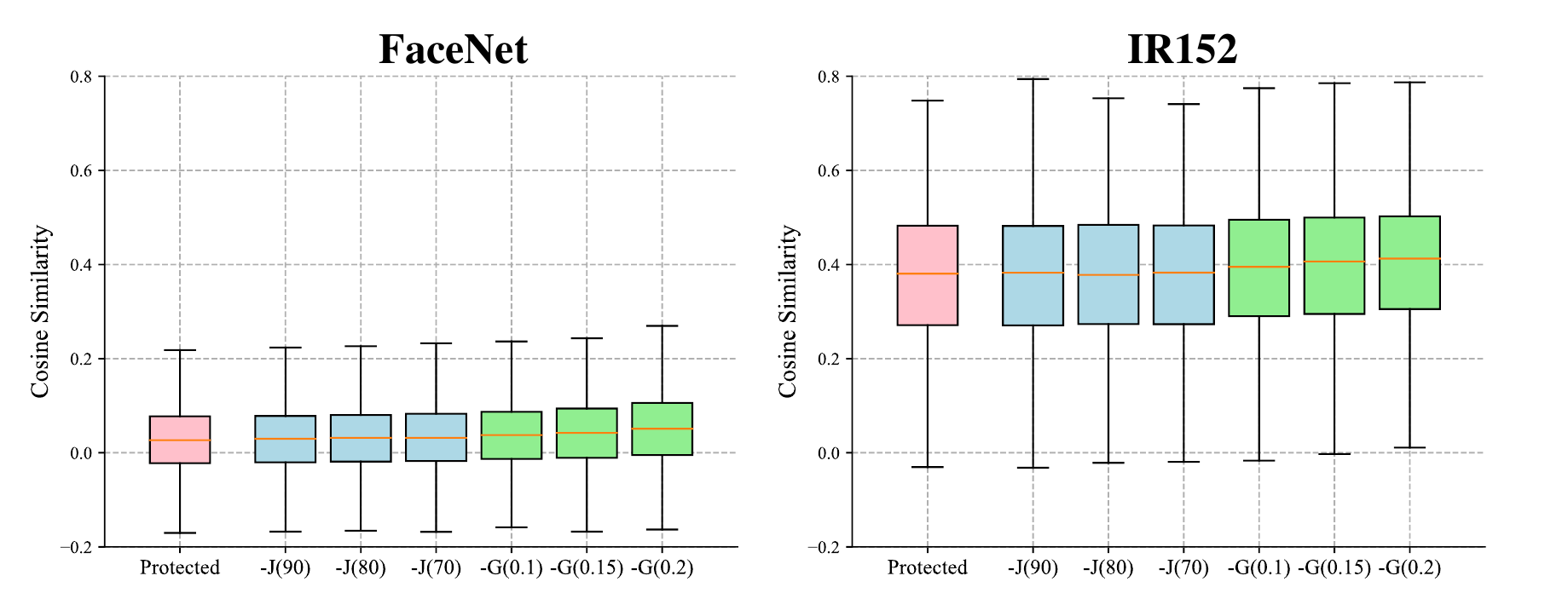}
	\caption{Identity similarity against JPEG  -J($Q$) and Gaussian  -G($\sigma$). Our method is robust enough against noise.}
	\label{my_robust}
\end{figure}

It is worth emphasizing that in real world scenarios, the protected image obtained by an adversary is never the same as the unprotected face it holds. The protected image and unprotected image are just different face images belonging to the same identity, which would further reduce the identity similarity and increase PSR.

\textit{2) PerceptFace exhibits satisfactory robustness to common noise and online social networks (OSNs). }

\textbf{ Against Noise.} Fig. \ref{my_robust} illustrates the similarity change under JPEG compression and Gaussian noise processing. We set the quality factor for JPEG compression to $Q = 70, 80, 90$ and  the standard deviation of Gaussian noise to $\sigma=0.1, 0.15, 0.2$. It can be observed that the noise introduced by these operations does not significantly change the identity similarity of the protected results, making the impact on privacy protection  weak. Therefore, PerceptFace is robust to common noise.

\begin{table}[!t]
	\begin{center}
		\caption{PSRs against OSNs in IMDB-WIKI.}
		\label{OSN_robustness}
		\scalebox{1}{
			\begin{tabular}{ccccccc}
				\toprule	
				&Facebook&Instagram&WeChat&QQ&Micro-blog\\
				\midrule
				
				Amazon&100.00\%&100.00\%&100\%&100\%&100\%\\
				Face++&100.00\%&100.00\%&100\%&100\%&100\%\\

				\bottomrule
			\end{tabular}
		}
	\end{center}	
\end{table}

\begin{figure}[!t]
	\centering
	\includegraphics[width=0.9\linewidth]{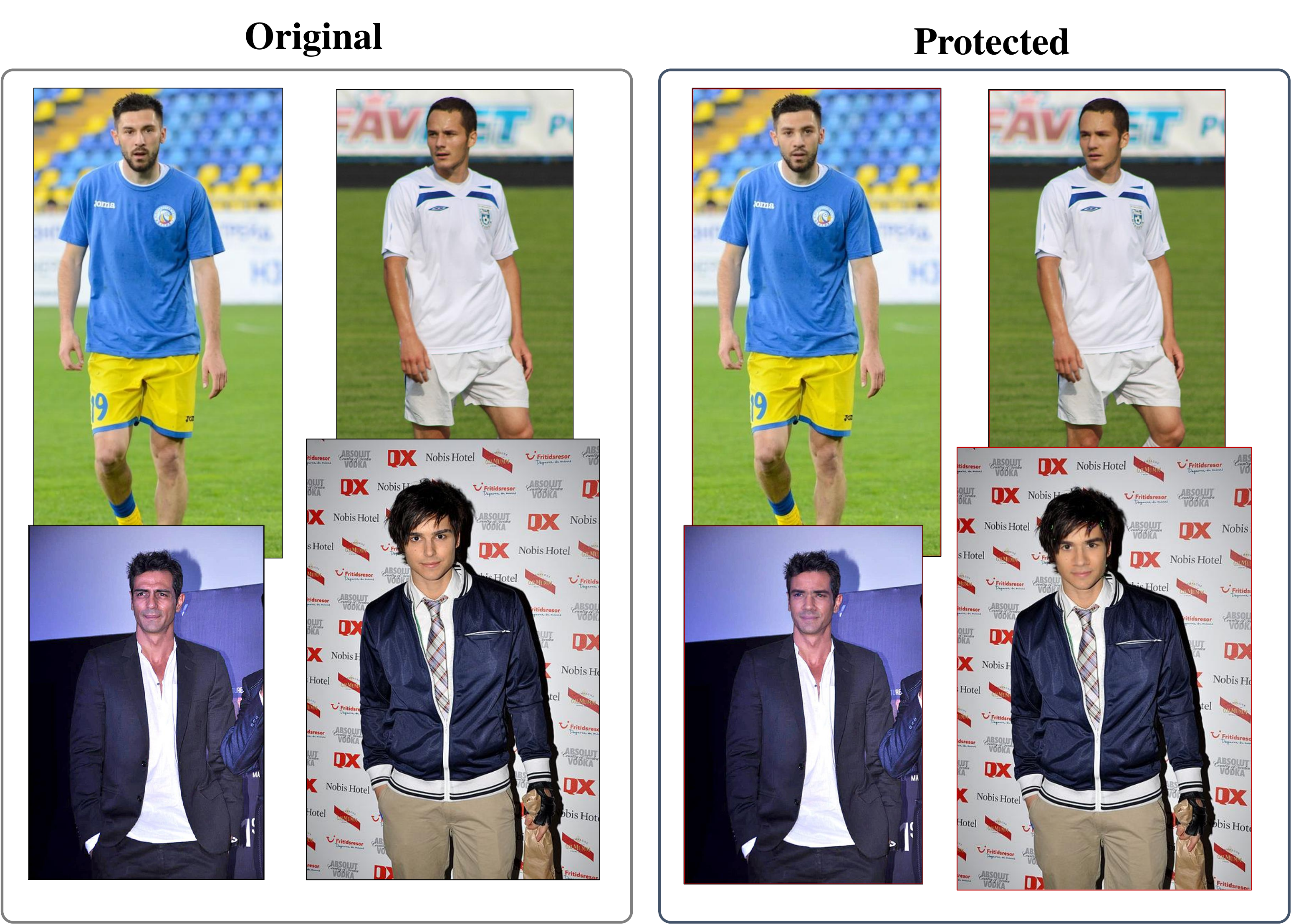}
	\caption{Examples with different sizes in IMDB-WIKI, simulating real photos instead of just human faces. }
	\label{OSN_photo}
\end{figure}

\begin{figure*}[!t]
	\centering
	\includegraphics[width=\linewidth]{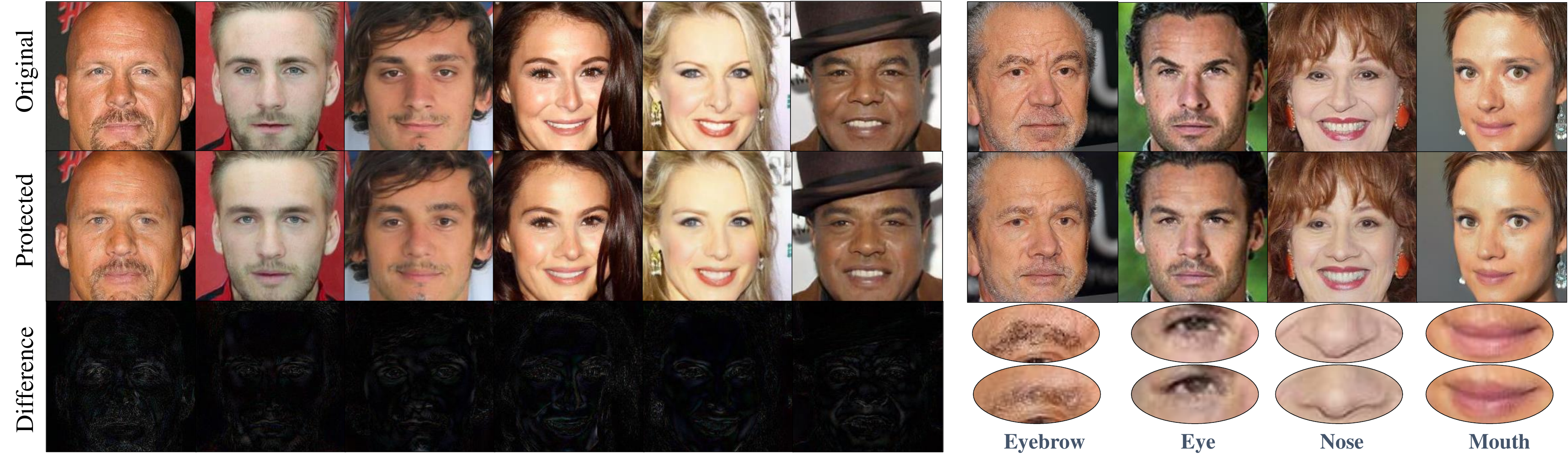}
	\caption{Pixel-level  and local region (the last four columns) visual  differences of our method.}
	\label{visual_difference}
\end{figure*}

\begin{table*}[!t]
	\begin{center}
		\caption{Image  similarity of different-level metrics.}
		\label{image_similarity}
		\scalebox{1}{		
			\begin{tabular}{ccccccc|ccccc}
				\toprule	
				&	&\multicolumn{5}{c|}{VGGFace2}&\multicolumn{5}{c}{CelebA-HQ}\\   
				\cmidrule(r){3-7}  \cmidrule(r){8-12}
				&&\multicolumn{2}{c}{Synthesis-based }&\multicolumn{2}{c}{Perturbation-based} 				&&\multicolumn{2}{c}{Synthesis-based }&\multicolumn{2}{c}{Perturbation-based}\\
				\cmidrule(r){3-4}  \cmidrule(r){5-6} \cmidrule(r){8-9}\cmidrule(r){10-11}
				
				&	&RiDDLE&Disguise&Fawkes&AMT-GAN&Ours	&RiDDLE&Disguise&Fawkes&AMT-GAN&Ours\\ 
				\midrule

				Perceptual Level&LPIPS$\downarrow$&0.343&\cellcolor{darkblue!20}0.100&\cellcolor{darkblue!50}0.033&0.114&\cellcolor{darkblue!35}0.059&0.283&\cellcolor{darkblue!20}0.087&\cellcolor{darkblue!50}0.023&0.088&\cellcolor{darkblue!35}0.058\\
				
				Structural level&SSIM$\uparrow$&0.499&0.737&\cellcolor{darkblue!50}0.976&\cellcolor{darkblue!35}0.850&\cellcolor{darkblue!20}0.826&0.493&0.764&\cellcolor{darkblue!50}0.980&\cellcolor{darkblue!35}0.863&\cellcolor{darkblue!20}0.842	\\

				Pixel level&L1$\downarrow$&0.102&\cellcolor{darkblue!20}0.060&\cellcolor{darkblue!50}0.009&0.081&\cellcolor{darkblue!35}0.032&0.097&\cellcolor{darkblue!20}0.055&\cellcolor{darkblue!50}0.008&0.069&\cellcolor{darkblue!35}0.031\\
				
				Pixel level	&RMSE$\downarrow$&0.142&\cellcolor{darkblue!20}0.089&\cellcolor{darkblue!50}0.016&0.103&\cellcolor{darkblue!35}0.048&0.136&\cellcolor{darkblue!20}0.082&\cellcolor{darkblue!50}0.015&0.091&\cellcolor{darkblue!35}0.047\\
				
				SNR level &PSNR$\uparrow$&7.096&\cellcolor{darkblue!20}21.179&\cellcolor{darkblue!50}35.818&20.150&\cellcolor{darkblue!35}26.562&17.420&\cellcolor{darkblue!20}21.857&\cellcolor{darkblue!50}36.813&21.197&\cellcolor{darkblue!35}26.664\\
				
				\bottomrule
			\end{tabular}
		}
	\end{center}	
\end{table*}

\textbf{ Against OSNs.}  In the real world, users often upload images of different sizes instead of the cropped fixed-size face images. For this reason, we select uncropped images for testing at IMDB-WIKI, where images containing the full body of the subject are more likely to be selected.  Considering that the image processing  in OSNs is not publicly available, we selected only 20 images to be manually uploaded and downloaded in OSNs, including Facebook, Instagram, WeChat, QQ, and Micro-blog. Fig.\ref{OSN_photo} shows the example results. We directly use face recognition APIs to detect faces and compare their identity similarity. As shown in Table \ref{OSN_robustness},  after the above OSNs processing, our method is still effective in protecting the identity. Therefore, PerceptFace is robust to OSNs.

Similar to RiDDLE and Disguise, PerceptFace makes change directly to the high-level semantics (identity). Benefiting from semantic robustness \cite{he2016deep}, the altered identity of the face in the protected image remains stably extracted under the influence of various noises, thus preventing the leakage of the original identity.

\subsection{Evaluation of  Utility Preservation}
PerceptFace applies subtle alteration
to only facial regions, which doesn’t affect the utility of photos (except facial identity). Therefore, the evaluation of utility preservation goal is only for identity perception.  Since human perception cannot be measured quantitatively, we use image similarity to approximate it. In addition, we also conduct user study for evaluating identity perception, which was done ethically.

\textbf{Ethical Considerations.} All participants provided informed consent for their data to be used in this research.  Given the minimal risk to individuals, the public nature
of the data, and the absence of sensitive information processing, our institutions consider this study exempt from IRB
review under relevant institutional guidelines
for low-risk research.

\textit{	1) PerceptFace exhibits significantly stronger image similarity performance than synthesis-based methods and  AMT-GAN (semantic level), and relatively low performance compared to Fawkes (pixel level). }

\textbf{High Image Similarity.} Table \ref{image_similarity} shows the results of various image similarity metrics. Based on different similarity levels, we chose a variety of metrics, including perceptual level (LPIPS), structural level (SSIM), pixel level (L1, RMSE), and signal-to-noise ratio (SNR) level (PSNR). Fawke achieves the best image similarity owing to the small addition of pixel-level noise, thus retaining the most utility. AMT-GAN achieves perturbation by changing the semantics (make-up), which alters image content to a greater extent. Thus, similar to Disguise, which only changes facial regions, AMT-GAN  only maintains a low image similarity. Our method further reduces the alteration of facial regions on the basis of Disguise, which enhances image similarity to retain more utility. This is because the designed perceptual loss limits the different variation of facial regions with different perceptual sensitivity.

\textbf{Low Visual Difference.} Fig.  \ref{visual_results_} shows a visual comparison of our method with baselines. It can be observed that the visual differences caused by our methods are imperceptible. Perturbation-based methods introduce visual unnaturalness, while synthesis-based methods have excessive visual content alterations. In Fig. \ref{visual_difference},  the results of the visual differences brought about by our method are further given. By careful observation, we can find weak facial contours, which can influence identity. Meanwhile, in Fig. \ref{visual_difference},  we carefully compare  faces before and after protection, and it can be found that the nose,  eyes and other identity-related areas  undergone different degrees of slight changes. However, in social photo sharing scenarios, friends simply do not compare the original face to notice such weak alterations. Therefore, our method achieves a very low visual difference.

\textit{ 2) PerceptFace maintains satisfactory identity perception and has an obvious preference  in user study. }

\textbf{Satisfactory Identity Perception.}  To simulate the prior knowledge of friends about users, we selected more famous celebrities for the evaluation, e.g., Cristiano Ronaldo. Specifically, we carefully selected and downloaded some photos containing them for 10 celebrities in the online Google search, which was done ethically. Medium resolution, frontal face, normal expression, and inclusion of upper body were more likely to be selected. For each celebrity, we constructed 2 test samples. Each test sample contains 3 photos of the celebrity and 6 photos of others, of which only one photo of the celebrity is protected. The task of the participant is to find all the photos of that celebrity in these 9 photos. \textit{As long as the protected photo is selected by the participant, the identity perception is successful}. If the participant  does not know the celebrity, he or she has the option of skipping the test sample. We collected a total of 600 answers from 30 participants, among which 36 answers indicated that they did not know the celebrity, so 564 valid answers were obtained after elimination. Statistical results show that 513 out of these 564 valid answers were able to include the corresponding celebrity, so the identity perception rate reached 90.96\%.

\textbf{Obvious Usage Preference.} We randomly selected 20 face images in CelebA-HQ and VGGFace2, where each image constitutes a test sample. Each test sample contains five protected faces (by baselines and our method, respectively), accompanied by a “no suitable option”.  We collected a total of 600 answers from 30 participants, and calculated the proportion of each method chosen  in Table \ref{Preference}.  No participants chose the synthetic-based methods because they significantly altered the perception of the  correct identity by human vision. Only a small number of responses chose the AMT-GAN, because it brought too much flashy make-up in most cases, which is hard for users to accept. Fawkes is also chosen by some participants who are able to tolerate poor visualization from noise. Benefiting from favorable visualization and slight alterations, our method received 64.5\% of  choices, which indicates an obvious preference of our method.

\begin{figure}[!t]
	\centering
	\includegraphics[width=0.8\linewidth]{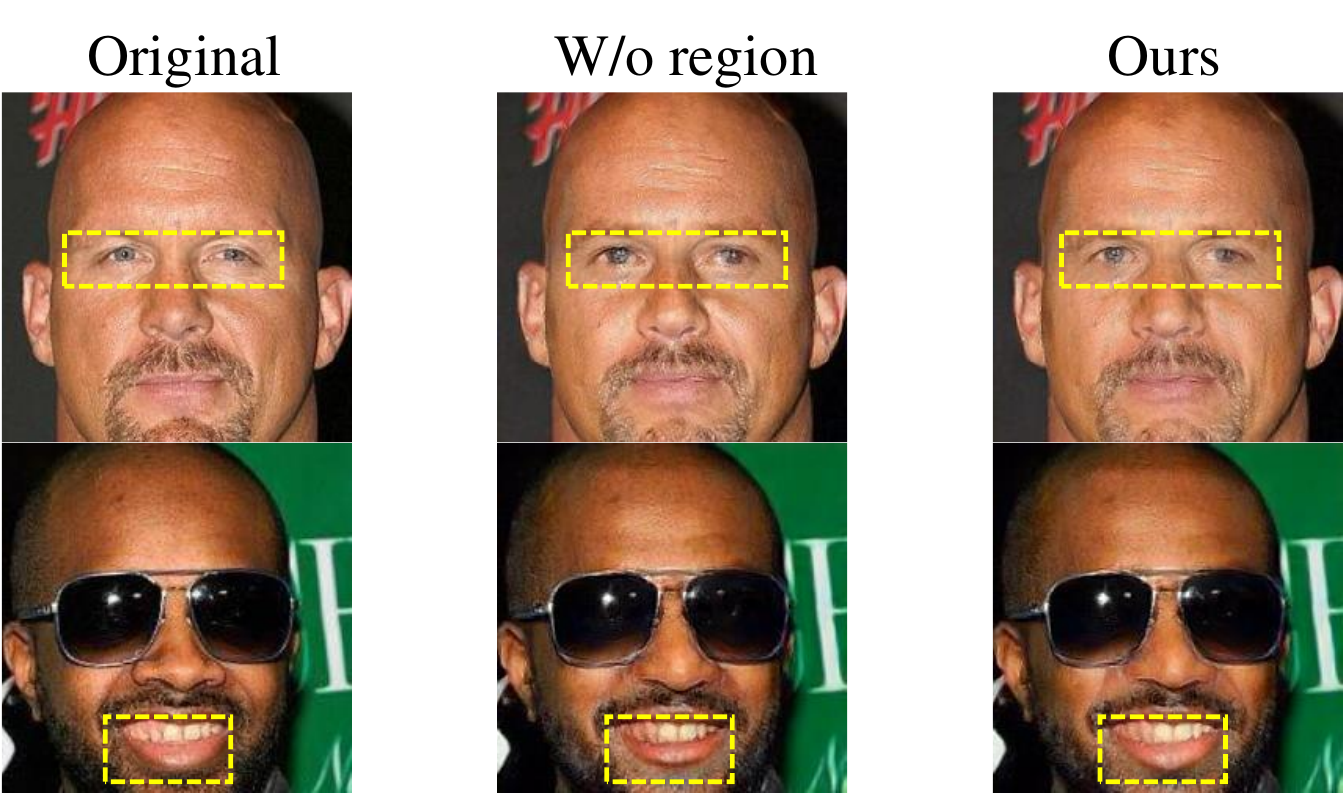}
	\caption{Ablation study, where "w/o region"  has a noticeable change in the eyes and lower lip, while ours is imperceptible.}
	\label{ablation}
\end{figure}

\begin{table}[!t]
	\begin{center}
		\caption{Usage preference in user study.}
		\label{Preference}
		\scalebox{1}{
			\begin{tabular}{ccccccc}
				\toprule	
				RiDDLE&Disguise&Fawkes&AMT-GAN&No option&Ours\\
				\midrule

				0.00\%&0.00\%&27.17\%&4.50\%&3.83\%&\textbf{64.50\%}\\
				
				\bottomrule
			\end{tabular}
		}
	\end{center}	
\end{table}

\subsection{Ablation Study}

To validate the effectiveness of the designed perceptual similarity loss, we remove the $\mathcal{L}_{region}$ from this loss (w/o region), i.e., only LPIPS was used.  \textbf{Qualitatively}, Fig. \ref{ablation} presents the visual results, where the significantly changed areas are outlined by the yellow dotted lines. It can be observed that the eyes of the first face become noticeably larger after the  protection of "w/o region", while ours also changes the eyes but in a relatively weak way. The second face becomes thin in the lower lip after the protection of "w/o region", while our change is imperceptible. As the findings of \cite{abudarham2016reverse}, human vision has higher sensitivity to eyes and lips, so our method can retain higher perceptual similarity than the "w/o region". \textbf{Quantitatively}, we directly calculate the change in the facial region using $\mathcal{L}_{region}$ and defined it as region\_MSE.
Table \ref{t_ablation} shows the results. Ours is able to provide higher region\_MSE (0.0636) at the expense of almost negligible LPIPS (0.0046),  reducing the alteration of highly sensitive regions.

\begin{table}[!t]
	\begin{center}
		\caption{Image change degree of ablation study.}
		\label{t_ablation}
		\scalebox{1}{
			\begin{tabular}{ccccccc}
				\toprule	
				&LPIPS$\downarrow$&Region\_MSE$\downarrow$\\
				\midrule
				W/o region&0.0548&0.1584\\
				
				Ours&0.0594(+0.0046)&0.0948(-0.0636)\\
				
				\bottomrule
			\end{tabular}
		}
	\end{center}	
	
\end{table}

\subsection{Limitation and Discussion}

We believe PerceptFace has the potential to become a practical anti-FR tool widely used on real OSNs. Of course, some limitations are worth discussing. 

\subsubsection{Applicable Photo Types} PerceptFace is mainly applicable to photos that contain the upper body of the subject and show him/her participating in an activity or event, which helps friends to perceive identity through contextual information of non-facial areas, e.g., a photo of the subject playing soccer, attending a party with friends, or blending in with the natural environment. In contrast, the tool is not applicable to photos that aim to represent the subject, e.g., selfies or portraits, because these photos usually have facial details as the main visual focus and lack rich contextual features, making the identity perception process more difficult.

\subsubsection{Performance Degradation from Distribution Shift} Faces in the real physical world have a different distribution than faces in the dataset, which degrades the performance of PerceptFace. Firstly, faces in datasets are aligned, whereas it is difficult to align and then protect faces in real-world photos. Secondly, faces in datasets are captured from early camera equipment, whereas the photos currently circulating on OSNs have higher resolution and more complex quality attributes with upgraded camera equipment. In the future, we consider leveraging domain adaptation, alignment-free modeling, and data augmentation techniques to address the performance degradation caused by distribution shift between dataset images and real-world photos.

\subsubsection{Identity Perception Quantization}  A user-friendly protection tool is suggested to provide quantifiable privacy and utility. Privacy can be represented by the difference in identity feature similarity. However, the utility provided by PerceptFace remains difficult to quantify, i.e., what is the face perceptual similarity before and after protection? In this paper, we only provide an optimization loss to improve the face perceptual similarity, and the complexity of human vision still makes it difficult to present a quantitative metric.

\subsubsection{Security Issues from Limited Diversity}  We further assume that an adversary can collect a large number of paired protected and unprotected faces. They can invalidate our method by training an en-decoder network. Achieving diversity in PerceptFace can be an effective solution but  is difficult. This is because identities with high perceptual similarity to a given face are rare. To improve security, we recommend that PerceptFace be securely managed and limited the frequency of its usage for a period of time.

\subsubsection{Failure Samples with Visible Nasal Distortion} 
\begin{figure}[!h]
	\centering
	\includegraphics[width=\linewidth]{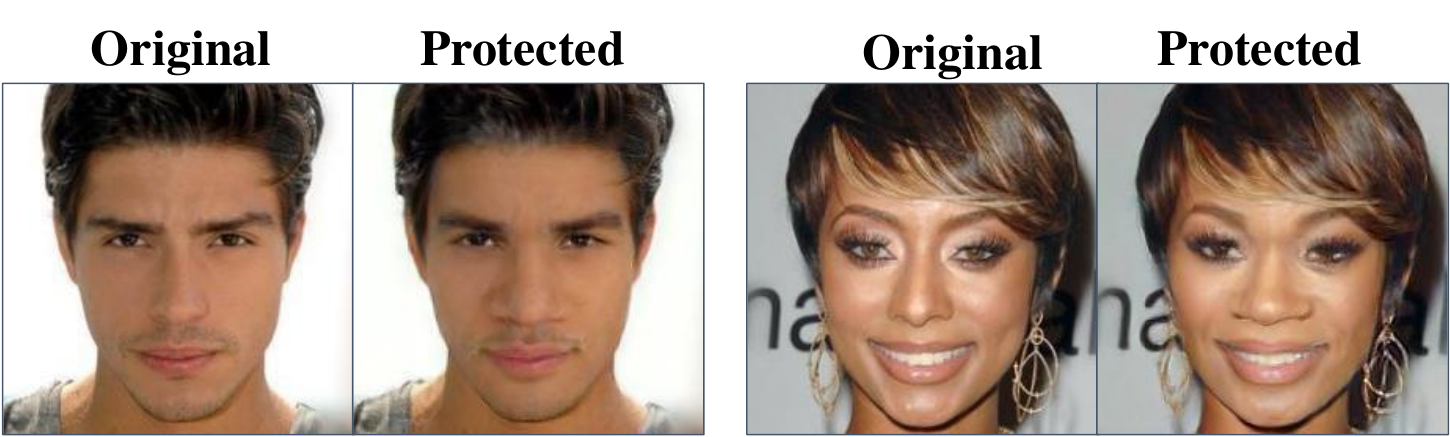}
	\caption{Failure samples with visible nasal distortion.}
	\label{Failure}
\end{figure}

Unlike perturbation-based methods that iteratively optimize a single photo, PerceptFace uses neural networks to uniformly model and protect photos. While this approach possesses higher processing efficiency, it is difficult to fine-tune it for each photo, and thus the desired visual effect may not be obtained on some photos. As shown in Fig. \ref{Failure}, we observe that the nose appears significantly enlarged in some samples. This is due to the relatively low perceptual sensitivity of the nose region, and the model tends to apply a large distortion to it to achieve effective changes in identity information. This phenomenon reflects the essential difference in the objective function design between PerceptFace and perturbation methods, as detailed in  Eq. \ref{eq1} and  Eq. \ref{eq2}. Nevertheless, PerceptFace enables users to intuitively perceive the degree of visual distortion, allowing them to make informed decisions about whether to share the photo. In contrast, although perturbation-based methods typically avoid noticeable distortion, their identity protection effectiveness is neither perceptible nor verifiable to users, making it difficult to establish sufficient trust.

\section{Conclusion}\label{sec5}

In this work, we explore a promising solution (synthesis-based methods)  toward the development of more practical anti-FR tools for subject faces in photos. Firstly, we reveal that perturbation-based methods provide just a false sense of privacy. Secondly, we present an insight that in most photo sharing scenarios, the recognition of subjects relies on identity perception rather than meticulous face  analysis by familiar persons.  Finally, based on the insight, we propose a novel synthesis-based method for subject face privacy, i.e., \texttt{PerceptFace}, which renders identity unextractable yet perceptible. Meanwhile, we also design a new face perceptual similarity loss which introduces perceptual sensitivity based on LPIPS. Sufficient experiments show that PerceptFace achieves a satisfactory balance between utility and privacy, which is expected to advance the real-world application of face privacy protection technology.

\bibliographystyle{IEEEtran}
\bibliography{bibliography}

	\begin{IEEEbiography}[{\includegraphics[width=1in,height=1.25in,clip,keepaspectratio]{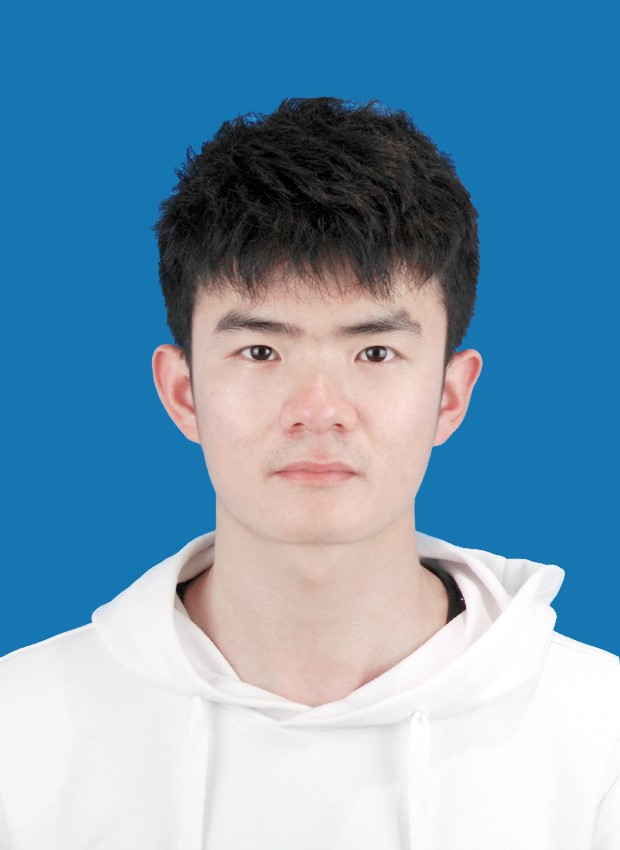}}]{Tao Wang}received the B.E. degree from the School of Computer and Information Technology, Anhui Normal University, Wuhu, China, in 2021 and the M.S. degree from the College of Computer Science and Technology, Nanjing University of Aeronautics and Astronautics, Nanjing, China, in 2024, where he is currently pursuing the Ph.D. degree. He has published several papers in top venues, e.g., IEEE TIFS, TDSC, ACM MM, CSUR. His current research interests include visual privacy, AIGC, adversarial perturbation, information theoretical privacy.
	\end{IEEEbiography}

	\begin{IEEEbiography}[{\includegraphics[width=1in,height=1.25in,clip,keepaspectratio]{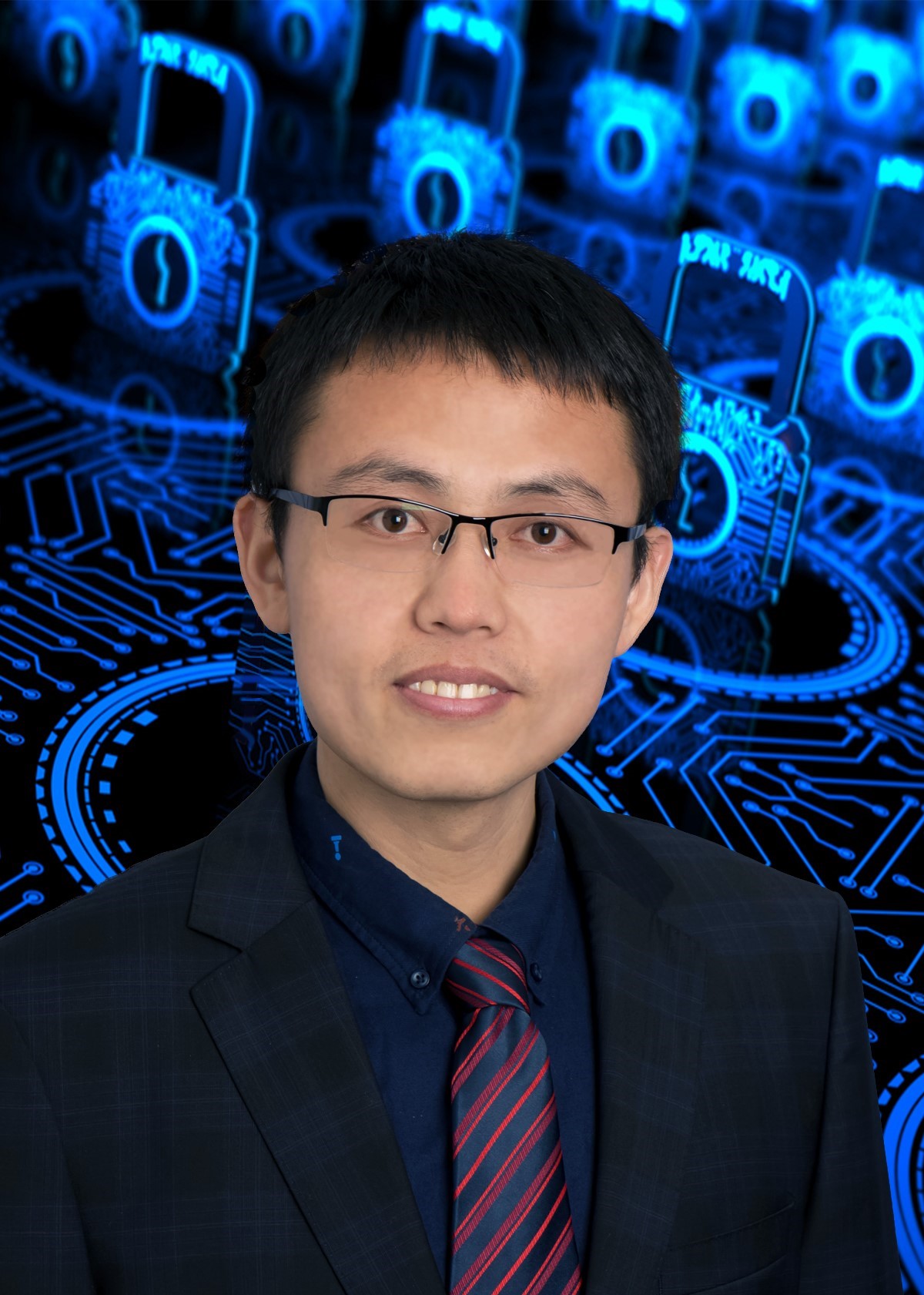}}]{Yushu Zhang} received the Ph.D. degree from the College of Computer Science, Chongqing University, Chongqing, China, in
	December 2014. He held various research positions
	with the City University of Hong Kong, Hong Kong;
	Southwest University, Chongqing; the University of
	Macau, Macau, China; and Deakin University, Gee
	long, VIC, Australia. He is currently a Professor in
	Jiangxi University of Finance and Economics.  He is an Associate Editor of  IEEE Transactions on Dependable and Secure Computing and IEEE Transactions on Network and Service Management, and Signal Processing. His research interests include multi
	media security, artificial intelligence security, and blockchain. \end{IEEEbiography}

	\begin{IEEEbiography}[{\includegraphics[width=1in,height=1.25in,clip,keepaspectratio]{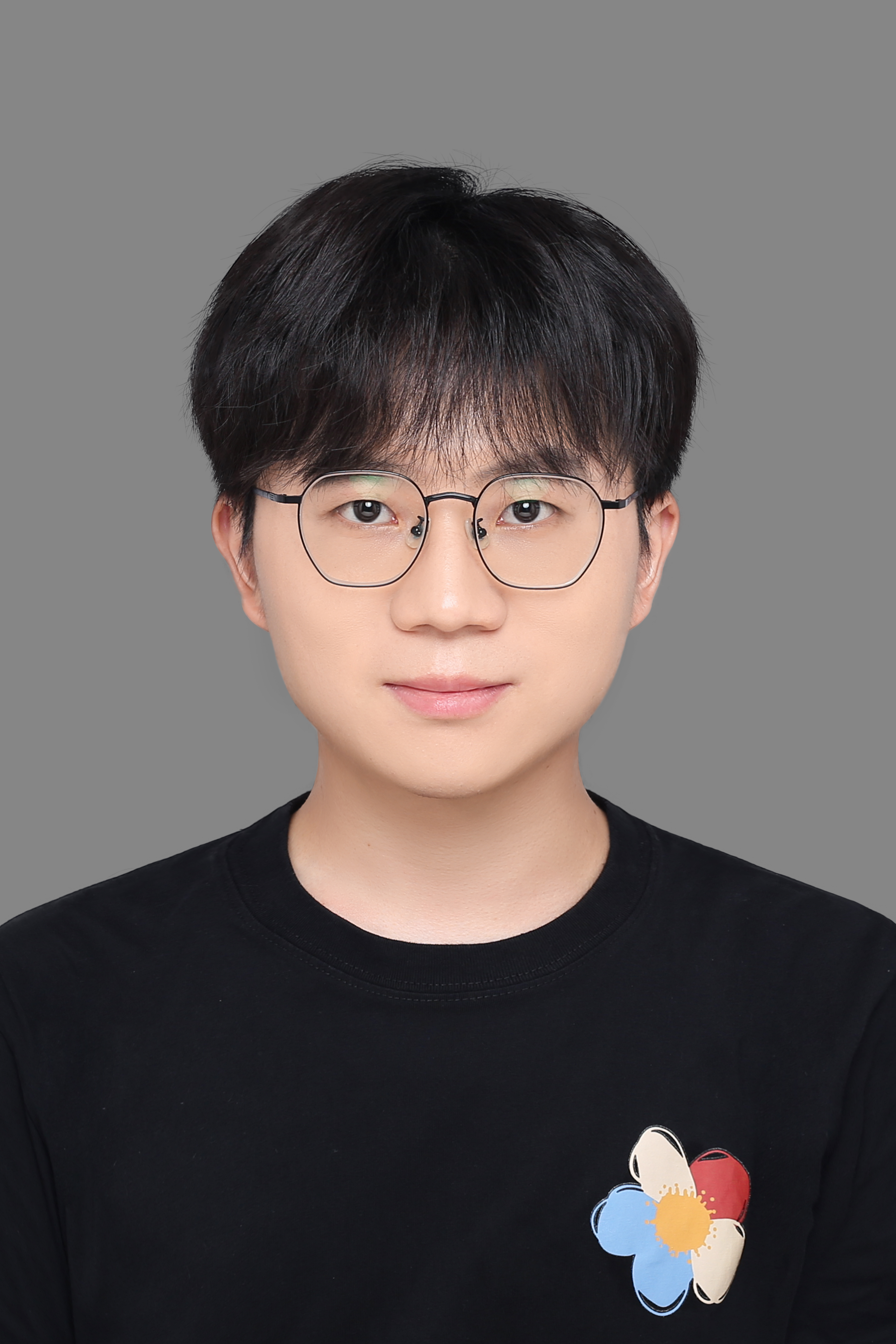}}]{Xiangli Xiao}   received the Ph.D. degree in cyberspace security from the College of Computer Science and Technology, Nanjing University of Aeronautics and Astronautics, Nanjing, China, in Oct. 2024. He is currently a Lecturer with the School of Computing and Artificial Intelligence, Jiangxi University of Finance and Economics, Nanchang, China. His current research interests include multimedia security, digital watermarking, and cloud computing security.
	\end{IEEEbiography}

    \begin{IEEEbiography}[{\includegraphics[width=1in,height=1.25in,clip,keepaspectratio]{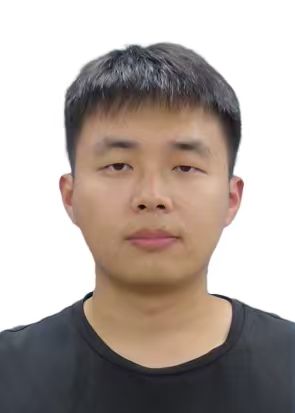}}]{Kun Xu}
received the B.E. and M.E. degrees from Anhui University of Science and Technology, Huainan, China, in 2020 and 2023, respectively. He is currently working toward the Ph.D. degree with the Nanjing University of Aeronautics and Astronautics, Nanjing, China. His research interests include generative model security.
\end{IEEEbiography}
	
	\begin{IEEEbiography}[{\includegraphics[width=1in,height=1.25in,clip,keepaspectratio]{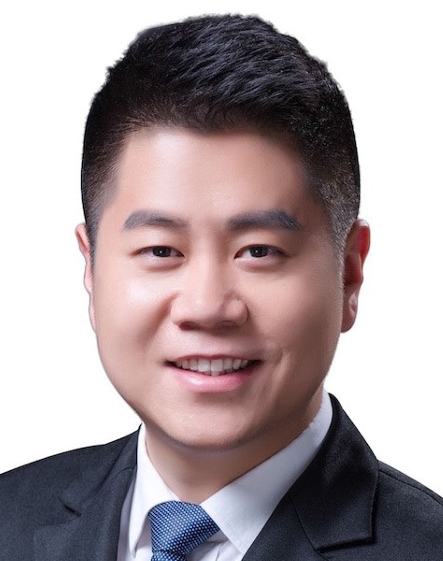}}]{Lin Yuan} received the B.Eng. degree in electronic science and technology from the University of Electronic Science and Technology of China (UESTC) in 2011 and the Ph.D. degree in electrical engineering from École Polytechnique Fédérale de Lausanne
 (EPFL), Switzerland, in 2017. He is currently an Associate Professor with the School of Cyber Security and Information Law, Chongqing University of Posts and Telecommunications. His research interests include image and video analysis, multimedia privacy protection, and media forensics.
	\end{IEEEbiography}

	\begin{IEEEbiography}[{\includegraphics[width=1in,height=1.25in,clip,keepaspectratio]{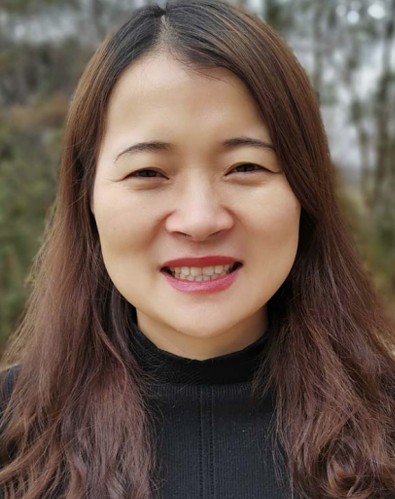}}]{Wenying Wen}  received the M.S. degree in computational mathematics from the Inner Mongolia University of Technology, Hohhot, China, in 2010, and the Ph.D. degree in computational mathematics from Chongqing University,Chongqing, China, in 2013. She is currently a Professor with  the School of Computing and Artificial Intelligence, Jiangxi University of Finance and Economics. Her research interests include image processing, and multimedia security, compressive sensing security, and blockchain.
\end{IEEEbiography}

\begin{IEEEbiography}[{\includegraphics[width=1in,height=1.25in,clip,keepaspectratio]{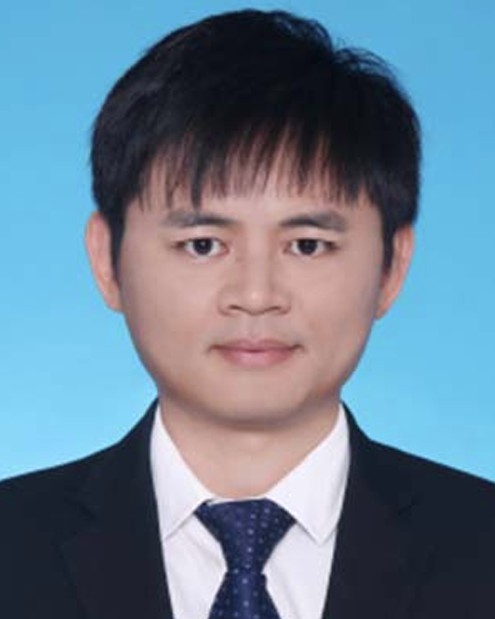}}]{Yuming Fang}  received the B.E. degree from Sichuan University, Chengdu, China, the M.S. degree from the Beijing University of Technology, Beijing, China, and the Ph.D. degree from Nanyang Technological University, Singapore. He is currently a Professor with the School of Information Management, Jiangxi University of Finance and Economics, Nanchang, China. His research interests include visual attention modeling, visual quality assessment, computer vision, and 3-D image/video processing.
\end{IEEEbiography}

\vfill

\end{document}